\documentclass[fleqn,10pt]{wlscirep}
\usepackage[utf8]{inputenc}
\usepackage[T1]{fontenc}
\usepackage{comment}
\usepackage{subfig}

\title{Urban Economic Fitness and Complexity from Patent Data}

\author[1,2,3,*]{Matteo Straccamore}
\author[4,1]{Matteo Bruno}
\author[3]{Bernardo Monechi}
\author[3,4,1,2]{Vittorio Loreto}
\affil[1]{Centro Ricerche Enrico Fermi (CREF), Rome, Italy}
\affil[2]{Sapienza Univ. of Rome, Physics Dept., Rome, Italy}
\affil[3]{SONY Computer Science Laboratories, Paris, France}
\affil[4]{SONY Computer Science Laboratories, Rome, Italy}

\affil[*]{matteo.straccamore@cref.it}


\begin{abstract}
Over the years, the growing availability of extensive datasets about registered patents allowed researchers to better understand technological innovation drivers. In this work, we investigate how the technological contents of patents characterise the development of metropolitan areas and how innovation is related to GDP per capita. Exploiting worldwide data from 1980 to 2014, and through network-based techniques that only use information about patents, we identify coherent distinguished groups of metropolitan areas, either clustered in the same geographical area or similar from an economic point of view. We also extend the concept of coherent diversification to patent production by showing how it represents a decisive factor in the economic growth of metropolitan areas. These results confirm a picture in which technological innovation can lead and steer the economic development of cities, opening, in this way, the possibility of adopting the tools introduced here to investigate the interplay between urban development and technological innovation.
\end{abstract}

\begin{document}

\flushbottom
\maketitle

    \thispagestyle{empty}

\section{Introduction}

Modern cities are at the centre of a passionate debate about their future. With over 55\% of the global population now living in urban areas, cities represent the core of the modern world. They are key for the production and diffusion of innovation~\cite{florida2017city,boschma2015relatedness} in many different sectors ranging from economy~\cite{jacobs2016economy} to science~\cite{leydesdorff2010mapping} and culture~\cite{pratt2008creative}. The ongoing pandemic has been imposing the hardest possible stress test on urban infrastructures and poses a real challenge in rethinking the role of cities, urban planning and policy decisions. While urbanisation keeps thriving~\cite{WinNT}, the challenge of understanding the development of cities to make them more sustainable and resilient becomes more and more crucial~\cite{kates2001sustainability,parris2003characterizing}. Therefore, it is of paramount importance to tackle urban areas' challenges by going beyond pure optimisation schemes and keeping a dynamic perspective. New tools are thus needed to understand and map the present and forecast how a change in the current conditions will affect and modify future scenarios.

Despite belonging to different geographical areas and socio-economic contexts, cities possess general features for economic development and urbanisation rates. For example, in~\cite{bettencourt2007growth}, authors show that many urban socio-economic indicators have a power-law correlation with the population size. In~\cite{hong2020universal}, the authors observe how individual cities recapitulate a common pathway where a transition to innovative economies takes place with a population of around 1.2 million. However, cities are ever-evolving systems where several changes and different growth paths are possible~\cite{bettencourt2020urban}. Technological innovation has been highlighted as the main driver for evolution and change in cities, and it is has been shown that complex economic activity flourish in large urban areas~\cite{balland2020complex}. In parallel, many studies recently focused on how innovation proceeds~\cite{tria2014dynamics,monechi2017waves,tacchella2020language}. In this paper, we focus on \emph{technological innovation}, and we investigate how the technological DNA of cities can affect their development and potential.

The adoption of patent data to monitor technological innovation is well established~\cite{frietsch2010value,griliches1998patents,leydesdorff2015patents}. For the past few decades, patent data have become a workhorse for the literature on technical change due mainly to the growing availability of data about patent documents~\cite{youn2015invention}. This ever-increasing data availability (e.g., PATSTAT, REGPAT and Google Patents~\cite{hall2001nber}) has facilitated and prompted researchers worldwide to investigate various questions regarding the patenting activity. For example, the nature of inventions, their network structure and their role in explaining the technological change~\cite{strumsky2011measuring,strumsky2012using,youn2015invention}. 

One of the characteristics of patent documents is the presence of codes associated with the claims contained in the patent applications. These codes mark the boundaries of the commercial exclusion rights demanded by inventors. Claims are classified based on the technological areas they impact according to existing classifications (e.g., the IPC classification~\cite{fall2003automated}) to allow the evaluation by patent offices. Mapping claims to classification codes allows localising patents and patent applications within the technology space. Many studies recently relied on network-based techniques to unfold the complex interplay among patents, technological codes and geographical reference areas. Network science techniques allowed to analyse economic activities of countries~\cite{pugliese2019unfolding}, regions~\cite{oneale2021structure,napolitano2018technology,dettmann2013determinants,tavassoli2014role,colombelli2014emergence}, cities~\cite{boschma2015relatedness,kogler2018patent,kogler2013mapping,balland2015technological} or firms~\cite{straccamore2022will,pugliese2019coherent}.\\

In the present work, we focus on cities to quantify the complexity of their technologies, correlating it with socio-economic indicators such as the GDP per capita. More precisely, we summarise our research questions as follows:\\
\textit{Which cities have the most advanced technological production?} We use the framework of Fitness and Complexity (FC)~\cite{tacchella2012new} to quantify the complexity of metropolitan areas and their technological endowment. Introduced initially and extensively adopted for countries' production/exports~\cite{tacchella2012new,zaccaria2014taxonomy}, the approach can easily be extended to any object pair, in this case, urban areas and technological codes.\\
\textit{Are cities able to diversify their production of patents, or do they tend to specialise in particular sectors?} In economics, FC has also been applied to sub-national scales, such as regions~\cite{operti2018dynamics,fritz2021economic} and firms, both at a country~\cite{bruno2018colombian} or global~\cite{laudati2022different} level. The study of bipartite economic systems at different scales revealed that to apply the FC framework, the economic agents need to have the capability to diversify to create global competition in the system. Otherwise, they will try to specialise and create a nested subsystem of entities specialising in the same products. In such a case, the analysis has to be restricted to subsystems for the FC method to capture the interplay among the economic agents. In this sense, the scale of the system is fundamental and regulates the interplay between competition and specialisation. We aim to understand whether metropolitan areas can compete globally or if they tend to specialise.\\
\textit{Are there clusters of cities with similar technological baskets?} Starting from a bipartite system of metropolitan areas - technology codes, we investigate the relations and similarities among metropolitan areas and uncover meaningful patterns in the evolution of their technological production. In bipartite systems, it is often important to understand the similarities between pairs of nodes of the same layer, to obtain a validated projection on a single layer~\cite{cimini2022meta}. We adopt this procedure to understand which metropolitan areas are more similar in the type of patents they produce and which patents are more likely to be produced together. 

The paper is organised as follows: in Section ~\ref{sec:data}, we describe the data used in this work and we go through our data cleaning procedure. In Section~\ref{sec:methods}, we introduce the methodologies used in our work, describing the details of the networks and measures we employed. In Section~\ref{sec:results}, we discuss the results showing how the network techniques can highlight non-trivial clusters of technologies and metropolitan areas, and how both the Fitness and the coherent diversification can drive a higher increase in the GDPpc of metropolitan areas. Finally, Section~\ref{sec:discussion} sums up our contributions and hints at future work needed to address questions arising from this study.

\section{Data}\label{sec:data}

\subsection*{Technology Codes}
Here, we shall adopt the PATSTAT database (\url{www.epo.org/searching-for-patents/business/patstat}) that provides information about patents and technology codes. The database contains approximately 100 million patents registered in about 100 Patent Offices. Each patent is associated with a code that uniquely identifies the patent and a certain number of associated technology codes. The WIPO (World International Patent Office) uses the IPC (International Patent Classification) standard~\cite{fall2003automated} to assign technology codes to each patent. IPC codes make a hierarchical classification based on six levels called digits, used to go into more and more detail about the technology used. The first digit represents the macro category: for example, the code Cxxxxx corresponds to the macro category "Chemistry; Metallurgy" and Hxxxxx to the macro category "Electricity"; considering the subsequent digits, we have, for instance, with C01xxx, the class "Inorganic Chemistry" and with C07xxx the class "Organic Chemistry".\\
After assigning a technology code to each patent, we use a database about cities (see next section) to match the unique patent identifier and its technology code to the corresponding city. To geolocalise the patents, we adopt the De Rassenfosse et al. database~\cite{de2019geocoding} that contains entries on 18 million patents from 1980 to 2014. Conveniently, in this database, the geographical information of patents is assigned to precise geographical coordinates. Thus, each patent has a unique identifier, a series of technology codes, and geographical coordinates identifying the corresponding city.

\subsection*{GDP of cities}
To obtain information on the GDP of cities and their evolution, we used the work of Kummu et al.~\cite{kummu2018gridded}. The authors constructed a worldwide GDP grid with a resolution of about five arc minutes for the 25 years 1990-2015. To compute the GDP per capita of each city or metropolitan area (MA) for each year in the data, we first download the boundaries from the Global Human Settlement Layer~\cite{Schiavina2019GHS}. Considering the GDP grid in one year, we compute the GPD per capita of a MA as the average of all the grid points within its boundaries. In Fig.3 in the Supplementary Information, we show the example of the grid of the Rome metropolitan area.

\subsection*{Data Cleaning Procedure}
To clean the data, the first step is to associate the technology codes of a patent with a specific city.
Once this preliminary operation is completed, it is possible to build the bipartite networks that will link cities to technology codes. We represent the bipartite networks through bi-adjacency rectangular matrices $V^y$ whose elements $V_{c,t}^y$ are integers indicating how many times a technology code $t$ appeared in different patents in a given city $c$ in year $y$.
In total, our network features $42912$ cities connected to $650$ technology codes (4-digit).
To reduce the difference between the two layers of the networks and reduce the noise in the system which is often due to the presence of very small cities, we aggregate the cities in the respective metropolitan areas (MAs). 
We select all cities within a metropolitan area (MA), and the technology codes associated with the metropolitan area will be the union of all the technology codes of the cities within it. The MAs present in the Global Human Settlement Layer~\cite{Schiavina2019GHS} are $8641$ and cover the entire world. However, most of these do not contain cities that have patents. The metropolitan areas producing patents are $2169$ and are distributed as shown in Figures 1 and 2 in the Supplementary Information.

We obtain a matrix $\textbf{V}^y$ for each year $y$ from $1980$ to $2014$, connecting $2169$ metropolitan areas $a$ and $650$ technology codes $t$. To avoid the fluctuations due to using only one year at a time as an interval, we decided to consider a window of $5$ years each time, summing the matrices in one window. In this paper, therefore, the matrix $\textbf{V}^y$ will refer to the time window from $y$ to $y+5$. The final database consists of $30$ 5-year window matrices $\textbf{V}^y$ ranging from window $1980-1984$ to $2010-2014$. Finally, we binarise the matrices $V$ applying a standard procedure in economic complexity to determine relevant producers/exporters of products (see Section \ref{subsection:RCA}). 

\begin{figure}
\centering
\subfloat[]
   {\includegraphics[width=.45\textwidth]{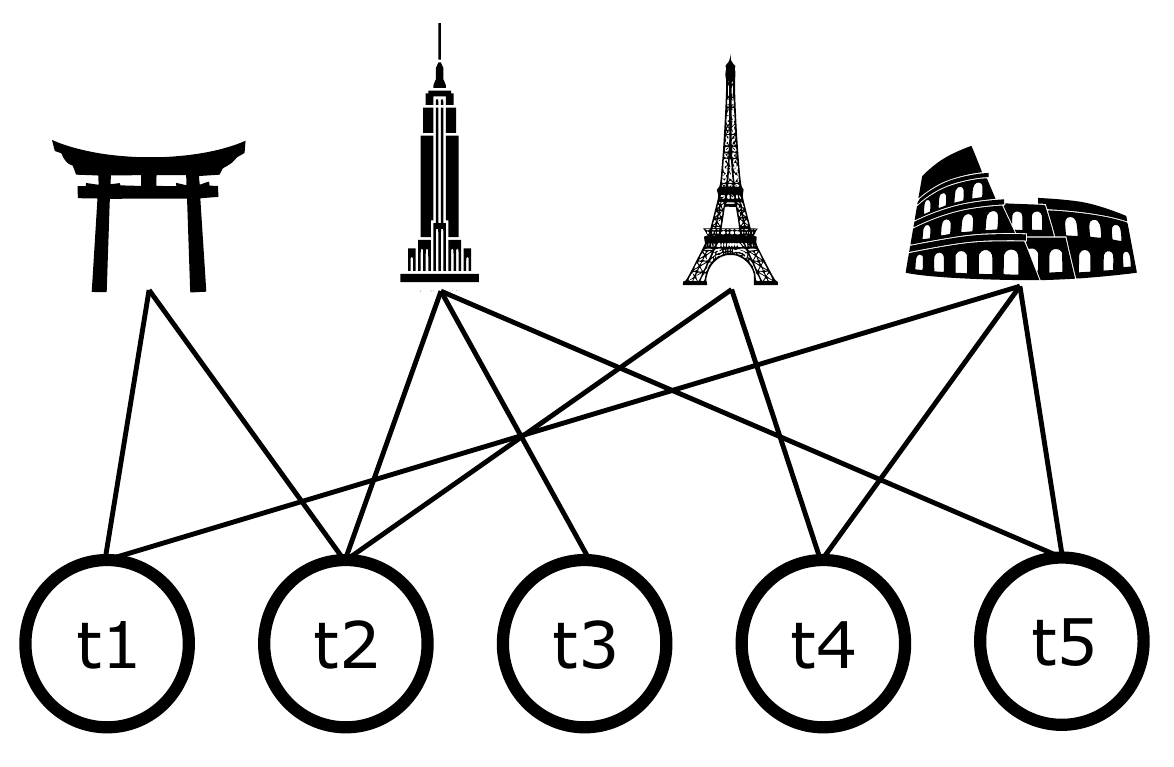}} \quad
\subfloat[]
   {\includegraphics[width=.45\textwidth]{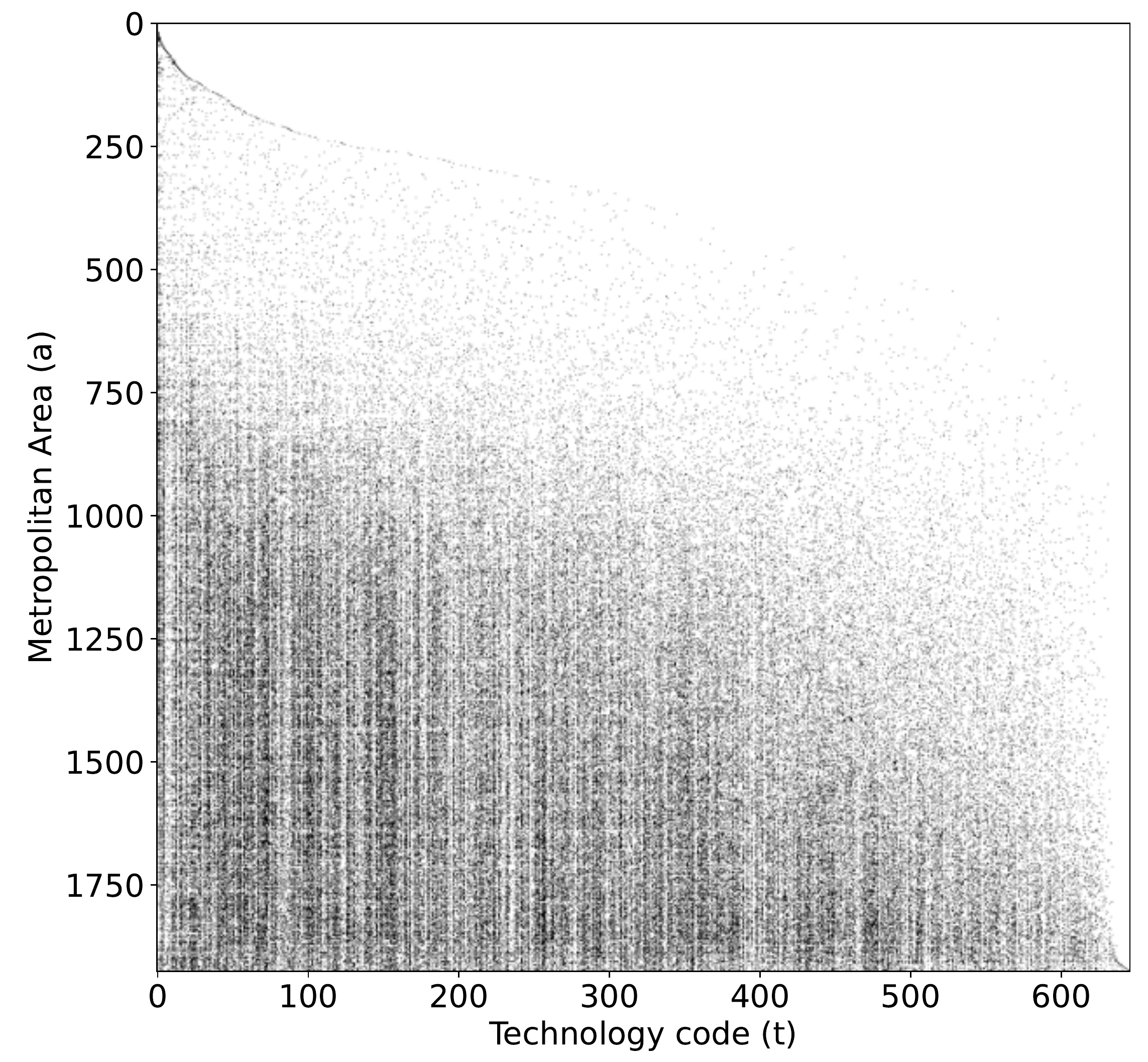}}
\caption{\textbf{Bipartite metropolitan areas - technology codes network.} \textbf{(a)}: Pictorial representation of the bipartite metropolitan areas-technology codes network. Each MA is connected to one or more technology sectors. \textbf{(b)}: Bipartite network adjacency matrix for the year 2000. A dark dot means that a given technology code is present in a patent made by a given MA.}
\label{fig:AM}
\end{figure}

\section{Methods}\label{sec:methods}

\subsection*{Revealed Comparative Advantage}\label{subsection:RCA}
To understand which metropolitan areas are relevant innovators of a specific technological sector, we apply the revealed comparative advantage (RCA)~\cite{balassa1965trade} binarisation strategy. RCA is a frequently used tool in the economic complexity literature~\cite{hidalgo2007product,zaccaria2014taxonomy,pugliese2019unfolding}. Considering a bipartite network of countries and products, RCA allows us to determine how competitive a country is in exporting a given product while also considering how many countries export that product. In our case, RCA reveals when the share of patents of some technology, $t$, introduced by a certain MA, $a$, is higher than the average share of the rest of the market, meaning that the metropolitan area focuses on the technology $t$ more than the number of technologies produced would suggest.

Considering the matrix $\textbf{V}^y$ for the year $y$, we define the RCA for the MA $a$ and the technology $t$ as:
\begin{equation*}
    RCA^y_{a,t} = \frac{V_{a,t}^y/\sum_{t'}{V_{a,t'}^y}}{\sum_{a'}{V_{a',t}^y}/\sum_{a',t'}{V_{a',t'}^y}},
\end{equation*}
\noindent
where the sums in the lhs run over all the technologies $t'$ and all the MAs $a$. A value $RCA_{a,t} \ge 1$ means that MA $a$ is significantly competitive in the technology field $t$. We use this threshold on the RCA values to obtain $30$ $\textbf{M}^y$ matrices, one for each $5$-year window:
\begin{equation*}
    M^y_{a,t} = \begin{cases}
    1\ \text{if}\ RCA^y_{a,t} \ge 1\\
    0\ \text{if}\ RCA^y_{a,t} < 1.
    \end{cases}
\end{equation*}
%
Notice that, in the following, we consider only having an average of at least one RCA $>1$ per year, reducing their number to $1211$. These $\textbf{M}^y$ matrices represent our final temporal bipartite network that links $1211$ MAs to $650$ technology codes.

\subsection*{Bipartite Networks}
A bipartite network is a network whose nodes represent two different kinds of entities, and only connections between nodes from different entities are allowed. Many systems in ecological and socio-economical environments, such as those studied in the present work, are easily described as bipartite since they involve interactions between two kinds of entities~\cite{straka2018ecology,bruno2018colombian}. For instance, the Internet can be modeled as a users-websites bipartite network, whose analysis can reveal sets and ranks of pages which will be more likely to be of interest for the user~\cite{zhou2007bipartite}. 
We can use the $\textbf{M}^y$ matrices as biadjacency matrices of MA - technology bipartite networks, connecting each MA with the technologies in which it is competitive. In figure \ref{fig:AM} we show a pictorial representation of this bipartite network and its biadjacency matrix $\textbf{M}^y$ for the year $y = 2000$.\\
Projecting the bipartite network on one of its layers, we can find non-trivial similarity patterns between MAs or technologies. However, the problem of finding the proper projection of a bipartite network into a monopartite one representing the similarities of nodes on one of its layers is well-known in the literature~\cite{saracco2017inferring,cimini2022meta,zhou2007bipartite,vasques2018degree}. In general, the goal is to find the representation of a monopartite network that best represents the bipartite one without taking too much information away from the latter. We decided to use the Bipartite Configuration Model (BiCM)~\cite{saracco2015randomizing,vallarano2021fast} to select the most significant nodes and links.


\subsubsection*{Bipartite Configuration Model (BiCM)}
One of the simplest ways to obtain a one-party projection from bipartite data is to count the number of links in common between two different entities belonging to the same layer. For example, using $\textbf{M}$ as the biadjacency matrix of a bipartite network between metropolitan areas $a$ and technologies $t$, 
counting the number of links in common between two different entities belonging to the same layer means computing:
\begin{equation*}
    A_{a_ia_j} = \sum_{t} M_{a_it}M_{a_jt},
\end{equation*}
where $A_{a_ia_j}$ is the adjacency monopartite projection matrix element of $\textbf{A}$ between elements $a_i$ and $a_j$. However, we note that a projection made in this way leads to a densely connected structure with a trivial topology.\\
To select the relevant nodes and links in our projected networks to avoid obtaining a too dense projection,
we use as a null model the Bipartite Configuration Model (BiCM)~\cite{saracco2015randomizing,saracco2017inferring,vallarano2021fast} which we compute by using the \textit{NEMtropy} Python package~\footnote{\url{github.com/nicoloval/NEMtropy}}. The BiCM belongs to the family of the Exponential Random Graphs, adapted to the case of bipartite networks. These models arise from the maximisation of the Shannon entropy of an ensemble of networks, in our case undirected binary bipartite networks $\textbf{M}$:
\begin{equation*}
    S = - \sum_{\textbf{M}\in \Omega}{P(\textbf{M})\ln{P(\textbf{M})}},
\end{equation*}
considering a set of constraints $\textbf{C}(M)$. $P(\textbf{M})$ is the probability of a specific bipartite network $\textbf{M}$.\\
The probability distribution maximising the entropy is the exponential distribution:
\begin{equation}
    P(\textbf{M}|\Vec{\lambda}) = \frac{e^{-H(\textbf{M},\Vec{\lambda}})}{Z(\Vec{\lambda})},
    \label{eq:P1}
\end{equation}
where $H(\textbf{M},\Vec{\lambda}) = \Vec{\lambda} \cdot \textbf{C}(M)$ is the Hamiltonian imposing the Lagrangian multipliers.\\
Two sets of constraints are imposed in the BiCM, one for each layer. Specifically, the node degrees are fixed, namely ubiquity $\Vec{u}(\textbf{M})$ for each technology code and diversification $\Vec{d}(\textbf{M})$ for MAs, in our case. The mean values of the node degrees must be tuned to match these quantities.
Then we obtain the Hamiltonian $H$:
\begin{equation*}
    H(\textbf{M},\Vec{\lambda}) = \Vec{\alpha}\cdot\Vec{d}(\textbf{M}) + \Vec{\beta}\cdot\Vec{u}(\textbf{M}).
\end{equation*}
Imposing the previous constraints together with the normalisation condition $\sum_{\textbf{M}\in\Omega}{P(\textbf{M})}=1$, we can write Eq. \ref{eq:P1} as:
\begin{equation*}
    P(\textbf{M}|\Vec{\lambda}) = \frac{e^{-\Vec{\alpha}\cdot\Vec{d}(\textbf{M}) - \Vec{\beta}\cdot\Vec{u}(\textbf{M})}}
    {\sum_{\textbf{M}}{e^{-\Vec{\alpha}\cdot\Vec{d}(\textbf{M}) - \Vec{\beta}\cdot\Vec{u}(\textbf{M})}}}.
\end{equation*}
Since constraints have been imposed on the mean values of the node degrees, the previous equation can be decomposed into the product of the probability distributions of a single link:
\begin{equation*}
    P(\textbf{M}|\Vec{\lambda}) =
    \prod_a\prod_t{p_{at}^{M_{at}}\left(1-p_{at}\right)^{1-M_{at}}}
\end{equation*}
where $p_{at} = \frac{x_ay_t}{1+x_ay_t}$ is the probability of the link between the MA $a$ and the technological code $t$, $x_a = e^{-\alpha_a}$ and $y_t = e^{-\beta_t}$.
To estimate the unknown parameters we have to maximise the log-likelihood $\mathcal{L}(\Vec{x},\Vec{y}) = \ln{P(\textbf{M}|\Vec{x},\Vec{y})}$, i.e. solving the system:
\begin{equation*}
    \Vec{\Delta}\mathcal{L}(\Vec{x},\Vec{y}) = 0
    \longrightarrow
    \begin{cases}
    d_a(\textbf{M}) = \sum_t{\frac{x_ay_t}{1+x_ay_t}}\  \forall\ a\\
    u_t(\textbf{M}) = \sum_a{\frac{x_ay_t}{1+x_ay_t}}\  \forall\ t
    \end{cases}
\end{equation*}
with $d_a(\textbf{M}) = d_a^*$ and $u_t(\textbf{M}) = u_t^*$ representing the observed quantities.\\

After we obtain the link probabilities of the model, we use them to compute how unexpected is the number of common neighbours of two nodes of the same layer. Given that, by construction, the links of the model are independent random variables, the probability of sharing a technology for two MAs is $P(V^{t}_{aa'} = 1)=p_{at}p_{a't}$, and the total number of technologies they share will be $V_{aa'} = \sum_t m_{at}m_{a't}$ . Thus, we can compute a p-value for the number of common neighbours observed for two nodes of the same layer, which reads:
\begin{equation}
    p\text{-value}_{aa'} = P(V_{aa'} > V_{aa'}^*)
\end{equation}
where $V_{aa'}^*$ is the number of common neighbours between nodes $a$ and $a'$ in the observed network. Note that the random variable $V_{aa'}$ is a Poisson-Binomial, i.e. a sum of independent Bernoulli random variables of different parameters, which is hard to evaluate when the number of different Bernoulli is large, we actually approximate this by substituting a Poisson variable with the same mean, as it has been done in previous works.

After applying this procedure to each pair of nodes, we obtain as output a p-value matrix of the same size as the adjacency matrix $\textbf{M}$ of the starting bipartite network. As a final step, we have to decide which of these $p$-values are significant and which are not. To assess the link significance, we use the False Discovery Rate test~\cite{benjamini1995controlling}: let us assume that we have $N$ hypotheses, each characterised by its p-value. The FDR first sorts these $N$ p-values as $p\text{-value}_1$,...,$p\text{-value}_N$, and then identifies the integer $I$ such that:
\begin{equation}
    p\text{-value}_I \le \frac{I\alpha}{N}
    \label{eq:pv}
\end{equation}
where $\alpha$ is the arbitrarily defined single-test significance level. We use $\alpha = 0.01$ for the projection onto the technology layer, and $\alpha = 0.1$ for the MA one. Note that in this case, $\alpha$ will be the statistical significance of the whole validated network, while for the single links their significance will be much lower. Finally, all hypotheses with p-value lower or equal than $p\text{-value}_I$ will be rejected, i.e. the link will be validated in the projected network.
In our case, for instance in the case of the projection on the technologies' layer, the number of hypotheses is the number of possible links in the projection $\binom{N_{t}}{2}$ and Eq.~\ref{eq:pv} becomes:
\begin{equation*}
    p\text{-value}_I \le \frac{I\alpha}{\binom{N_{t}}{2}}.
\end{equation*}
Ordering the coefficients $\binom{N_{t}}{2} p\text{-value}_(V_{tt'})$ and retaining only the links between pairs of nodes $t, t'$ such that $p\text{-value}(V^*_{tt'})\le p\text{-value}_I$ yields our projection. 

Let us remark that the projection obtained via the procedure just described only keeps links that are highly significant with respect to the degree of the nodes, unveiling hidden strong similarities.

\subsubsection*{Modularity and Community detection}
We are interested in finding relevant communities of MAs or technologies to visualise better which nodes in the two layers are highly interconnected. To this end, we adopted the Louvain method introduced by Blondel et al.~\cite{blondel2008fast}, which relies on finding a partition that maximises the modularity.
We also vary the Resolution~\cite{lambiotte2008laplacian} to find communities at different scales.

\subsection*{Fitness and Complexity algorithms}
The Fitness and Complexity (FC) framework~\cite{tacchella2012new}, introduced in $2012$, provides a way to quantify the competitiveness (Fitness) of the economy of a country.
Here, we adopt it to quantify the Fitness of metropolitan areas considering only patent data. The idea is to define an iterative process linking and combining the Fitness of a MA, $F_a$, with the Complexity of a specific technology, $Q_t$. The iterations to find these quantities are defined as:
\begin{equation}
    \begin{cases}
    \tilde{F}_a^{n+1} = \sum_t{M_{at}Q^{n}_t}\\ \tilde{Q}_t^{n+1} = \frac{1}{\sum_a{\frac{M_{at}}{F^{n}_a}}}
    \end{cases}
\label{eq:FCalg}
\end{equation}
where for each step $n$ the quantities are normalised as:
\begin{equation}
    \begin{cases}
    F^n_a = \frac{\tilde{F}_a^n}{\left<\tilde{F}\right>_a}\\ 
    Q_t^n = \frac{\tilde{Q}_t^n}{\left<\tilde{Q}\right>_t}
    \end{cases}
\end{equation}
and initial conditions $Q_t^{(0)} = 1\ \forall t$, $F_a^{(0)} = 1\ \forall a$. In~\cite{pugliese2016convergence} the convergence of the algorithm is studied in detail. In our case, we compute $F^y_a$ and $C^y_t$ for each 5-years window $y$ starting from the biadjacency matrices $M^y_{at}$. We stop the iteration when the Fitness ranking of MAs does not change anymore.
The rationale behind the whole process is as follows. A technology made in an already developed MA carries little information about the complexity of the technology itself because developed metropolitan areas produce a large part of the technologies. In contrast, a technology exported by an underdeveloped MA must require a low level of sophistication. Thus, it is possible to measure a MA's technological competitiveness given its technologies' complexity. A different approach should be taken instead to assess product quality. Fitness $F_a$ is proportional to the sum of technologies weighted by their complexity $Q_t$. Intuitively, the complexity of a technology is inversely proportional to the number of MAs that have implemented it. If a MA has high Fitness, this should reduce the burden of limiting the complexity of a technology, and MAs with low Fitness should contribute strongly to $Q_t$.\\
In recent studies, the authors of ~\cite{operti2018dynamics,sbardella2021behind} have shown that it is helpful to calculate the Fitness of sub-national actors using the complexity that comes from the national systems. This measure is called exogenous Fitness and overcomes the issue of the limited capabilities of sub-national entities, such as cities/MAs in our case. Thus, as for Fitness calculations, we enter the complexity obtained by considering global international patent data instead of calculating the complexity of a technology only on the MA subsample. We proceed in the same way by aggregating all the MAs of a country, i.e., summing all the rows of the MAs and running the FC algorithm. In other words, we compute $F^C$ and $Q^C$ relative to each country $c$ and technology $t$ through the formulas~\ref{eq:FCalg}, and then calculate the Fitness of the MAs through:
\begin{equation*}
    F^{MA}_a = \sum_t{M_{at}Q^C_t}.
\end{equation*}
For each time window, we calculate the Exogenous Fitness of all metropolitan areas and the complexity of each technology.

\subsection*{Coherent diversification}

The coherence of production and innovation diversification has been shown to be a significant driver of productivity~\cite{quatraro2010knowledge,kalapouti2015intra}. Thus, to better understand the nature of MAs' performance from their technology portfolio, we analyze their coherent diversification~\cite{pugliese2019coherent}. The underlying question is whether the accumulation of knowledge and capabilities associated with a coherent set of technologies leads MAs to experience more significant benefits in terms of GDPpc. Consistent diversification is defined as the
Coherence of the technology field $t$ with respect to the technology basket of the MA $a$:
\begin{equation}
    \gamma_{at}=\sum_{t' \neq t}{B_{tt'}M_{at'}}.
    \label{eq:coer1}
\end{equation}
where $B$ can be any matrix quantifying the similarities between pairs of technologies and $M$ is the usual adjacency matrix of a bipartite network between the layers of MAs and technologies. For each technological field, $t$, and each MA, a, one counts how many technologies $t^{\prime}$ adopted by a are connected with $t$, using $B_{tt^{\prime}}$ as a weight. If the technological portfolio of a is such that t is surrounded by 
a large number of strongly connected technologies owned by a, then $t$ will be very coherent to a, and $\gamma_{at}$ will be high. On the contrary, if $t$ belongs to a portion of the network of technologies far from the patenting activity of a, $\gamma_{at}$  will be low.  In our case, we use as $B$ matrix the projection represented in Fig.~\ref{fig:netw_tec}. Notice that $\gamma$ has the same dimensions as $M$, and the elements quantify how coherent a technology $t$ is to the technology basket of MA $a$. 

Finally, we can calculate the coherent technological diversification~\cite{pugliese2019coherent} of MA, $a$, as:
\begin{equation}
    \Gamma_a = \frac{\sum_{t}{M_{at}{\gamma_{at}}}}{d_a},
    \label{eq:coer2}
\end{equation}
where $d_a = \sum_t{M_{at}}$ is the diversification of MA, $a$.
The Coherence of technological diversification, $\Gamma_a$, of MA $a$ computes the average coherence $\gamma$ of the technologies in which $a$ is patenting.

\section{Results}\label{sec:results}

\subsection*{Networks projection}
%
To find a general network representation of our data for each year, we apply the Bipartite Configuration Model (BiCM) projection method (discussed in detail in the Methods section) to each $\textbf{M}^y$ matrix, one for each 5-years window (for both layers of technology codes and MAs) using the following steps:
\begin{itemize}
    \item For each 5-year sliding window, we calculate the BiCM projection (with the same parameters every year), which gives us the most relevant nodes and links;
    \item We merge all the projections for every year as follows. For instance, suppose code A00A is connected with A00B in the projection relative to $1980$, but code A00C does not appear in this network; suppose also that in the network of $1981$, A00A is connected with A00C, but A00B does not appear. In the merged network, we will have both a link between A00A and A00B and a link between A00A and A00C;
    \item We use weights: e.g., if A00A is connected with A00B in $1980$, $1981$ and $1982$, the relative link weight will be 3. We decided to do this to emphasise a relevant link between relevant nodes that lasts over time.
\end{itemize}
The resulting technology network was obtained by setting the BiCM parameter for the statistical validation of the projected networks to $\alpha = 0.01$ for every year. In contrast, the projected networks of MAs were obtained by setting the threshold $\alpha=0.1$, as explained in the Methods section.
The two networks have a density of $0.032$ and $0.012$ for technologies and MAs, respectively. The mean density of the starting bipartite ones in years is $0.124$. After these steps, we use the Louvain algorithm to identify communities, as discussed in the Methods section.
\begin{figure}[h!]
\centering
\subfloat[]
   {\includegraphics[width=.47\textwidth]{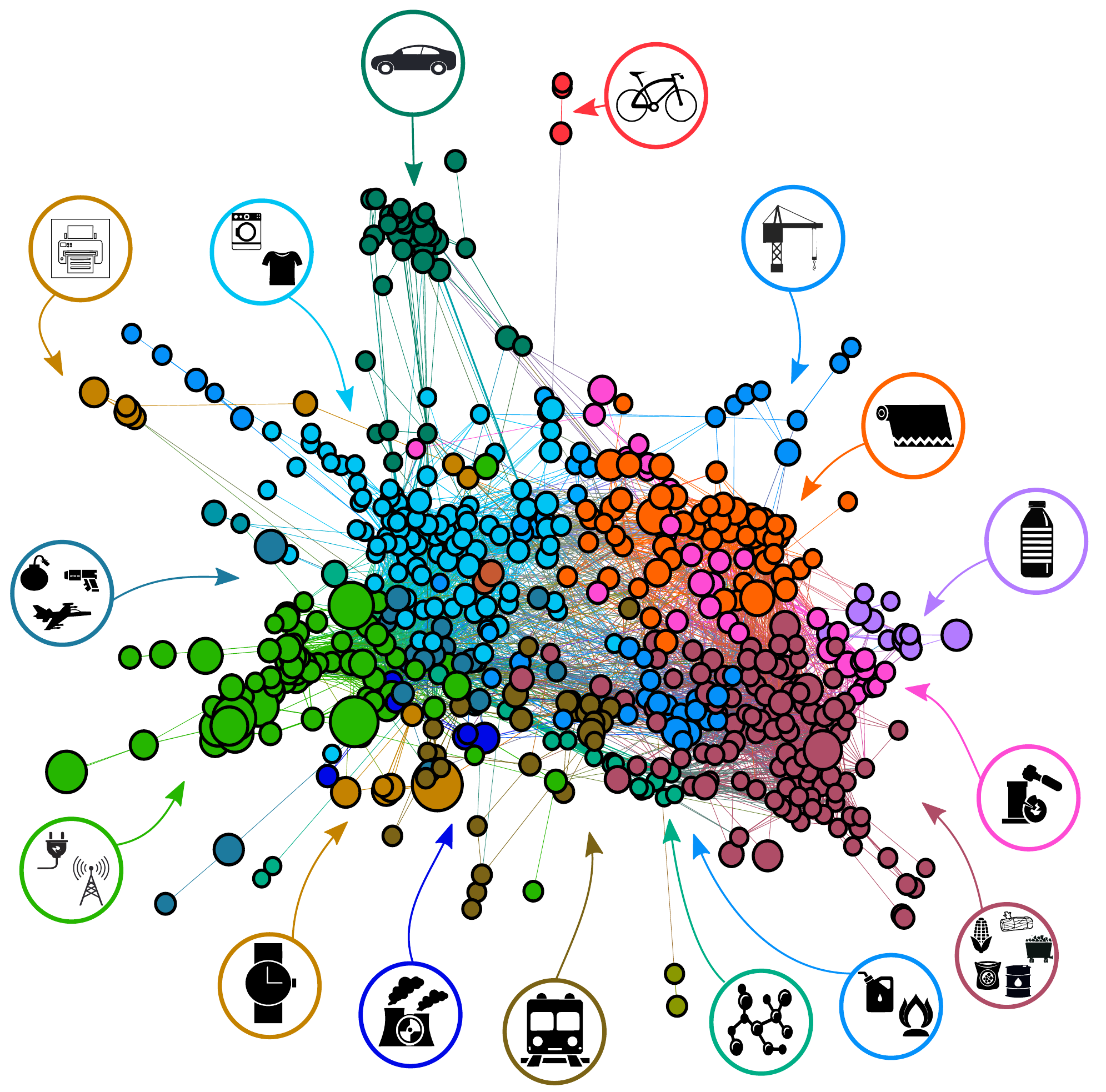}\label{fig:NTN}} \quad
\subfloat[]
   {\includegraphics[width=.47\textwidth]{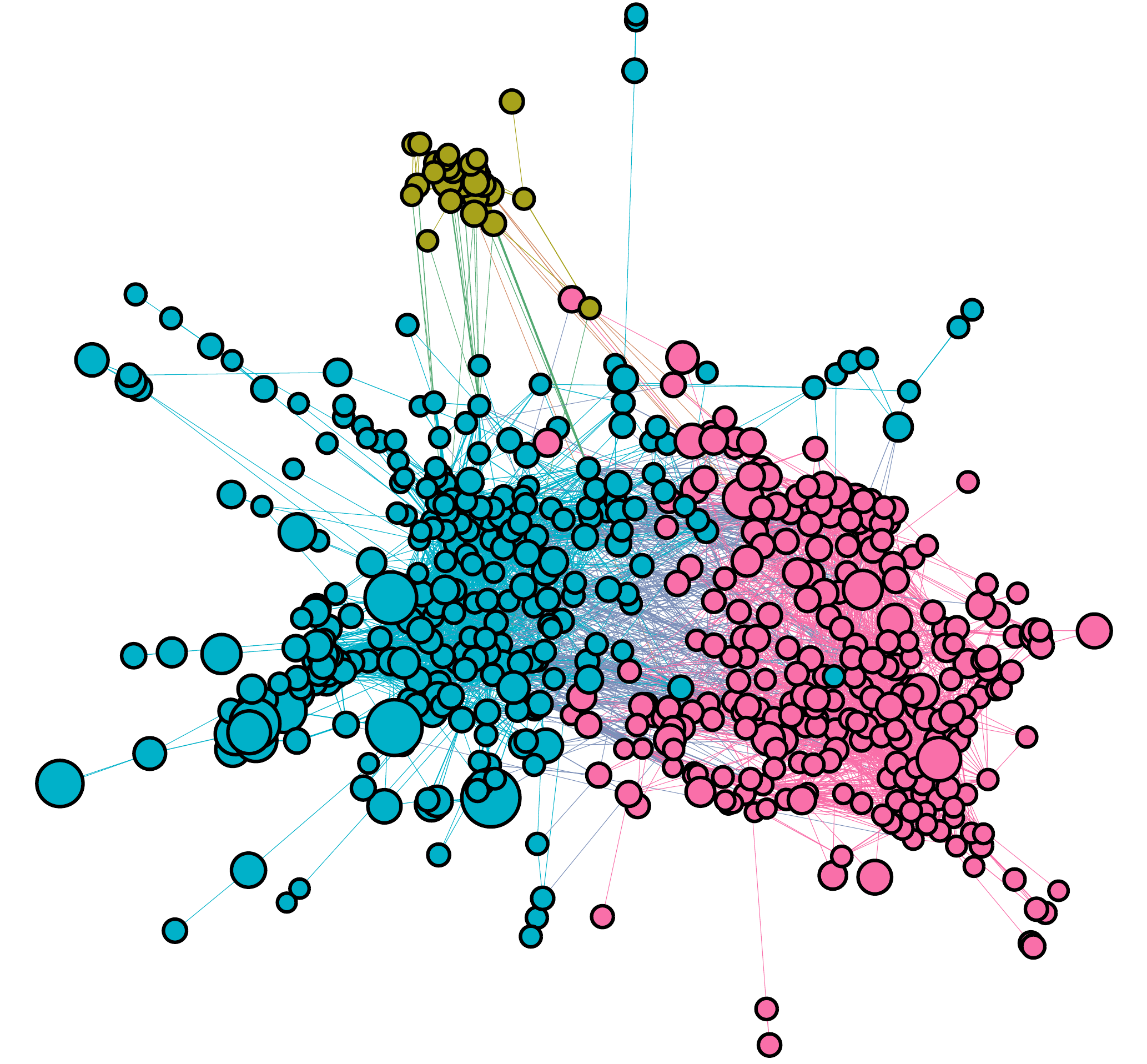}\label{fig:net}} \\
\subfloat[]
   {\includegraphics[width=.43\textwidth]{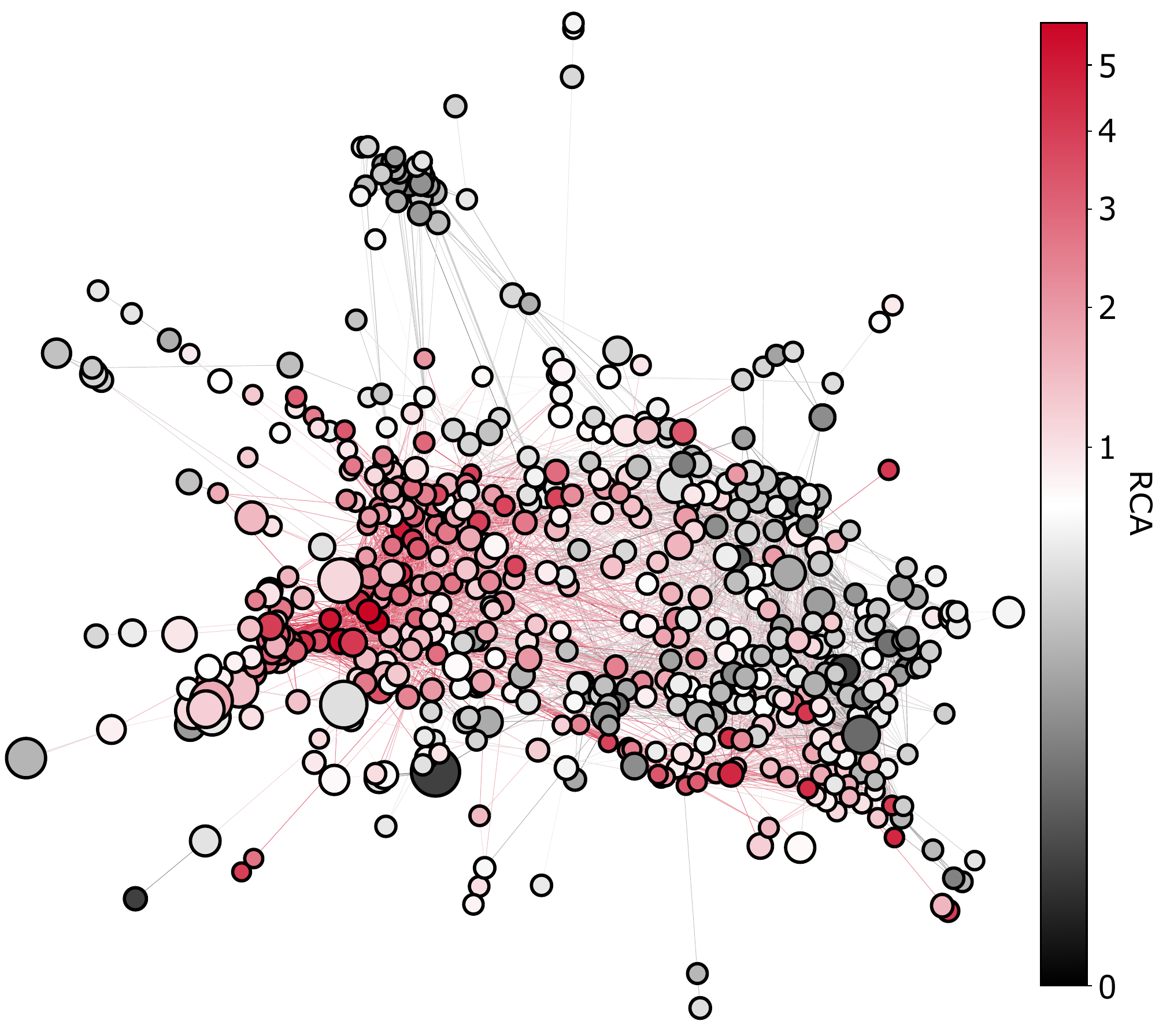}} \quad
\subfloat[]
   {\includegraphics[width=.43\textwidth]{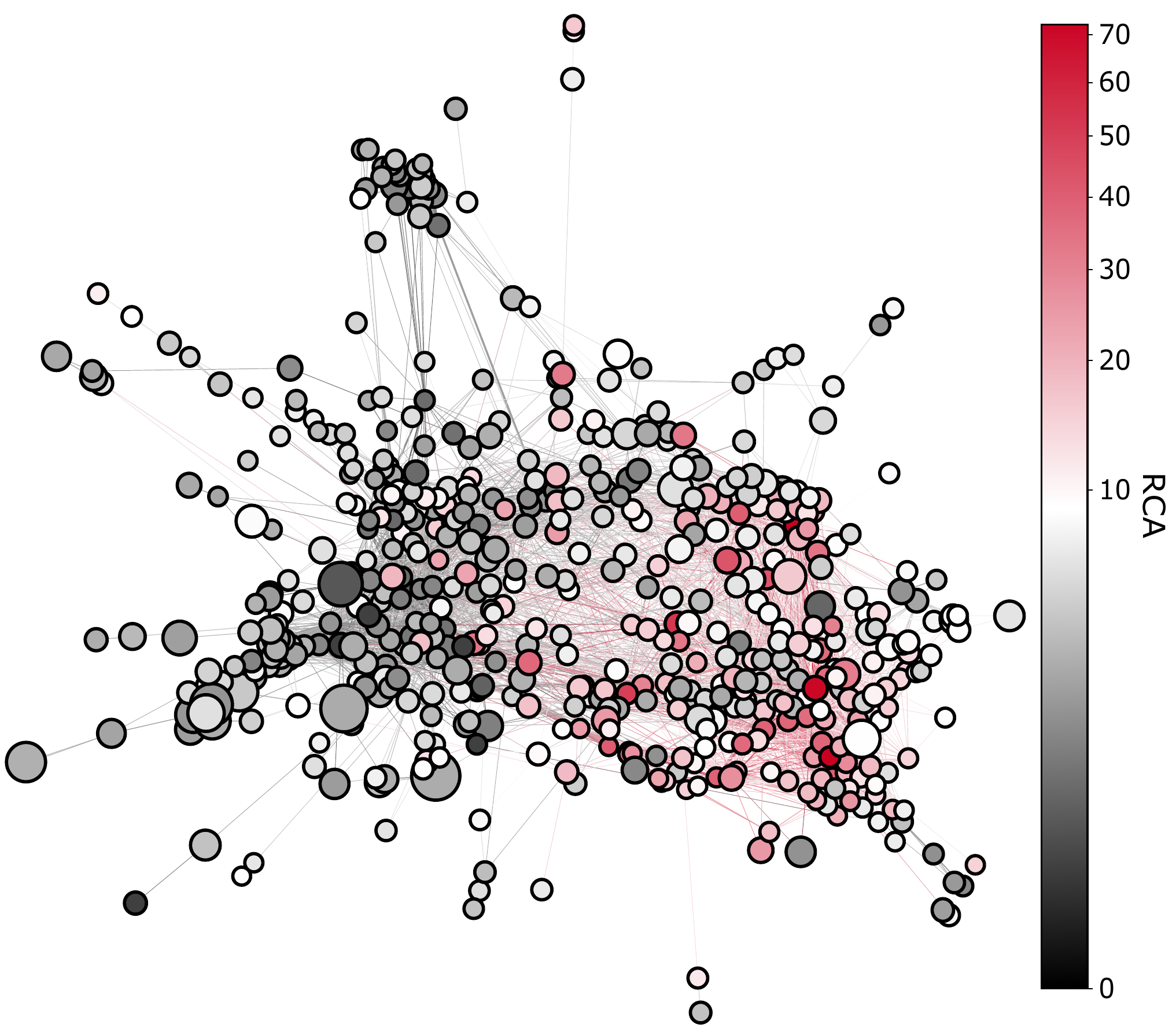}}
\caption{\textbf{Projected network of technologies.} Each node in these figures is a technology code. The size of the nodes is proportional to the complexity of the technology. \textbf{(a)}: clusters found by the community detection algorithm with a modularity value of $0.56$ and a density of $0.032$. We could identify the specific significant technologies for most clusters and represent them with corresponding icons in coloured circles. The Electricity and Information cluster (light green on the left) contains the most complex technologies. \textbf{(b)}: clusters found by the community detection algorithm in the technology network with higher resolution, i.e., considering larger communities. In this case, we could identify three regions. Olive green: this region contains technologies closely related to cars; light blue: this macro area contains clusters of technology sectors that we can classify as highly sophisticated technology sectors; pink: clusters related to manufacturing technology sectors. \textbf{(c)} and \textbf{(d)}: Projection on the technology network of average RCA values in the database years of New York (\textbf{(c)}) and Shanghai (\textbf{(d)}). The colour scale shows the RCA value of the metropolitan area in a particular technology: more red nodes (technologies) indicate a high RCA value of the MA for those specific technologies. We note how Shanghai has focused more on manufacturing technologies (pink region of \textbf{(b)}), while New York is strong in electricity and communications technologies (light blue region of \textbf{(b)}).}
\label{fig:netw_tec}
\end{figure}
\begin{figure}[h!]
    \centering
    \includegraphics[scale=.55]{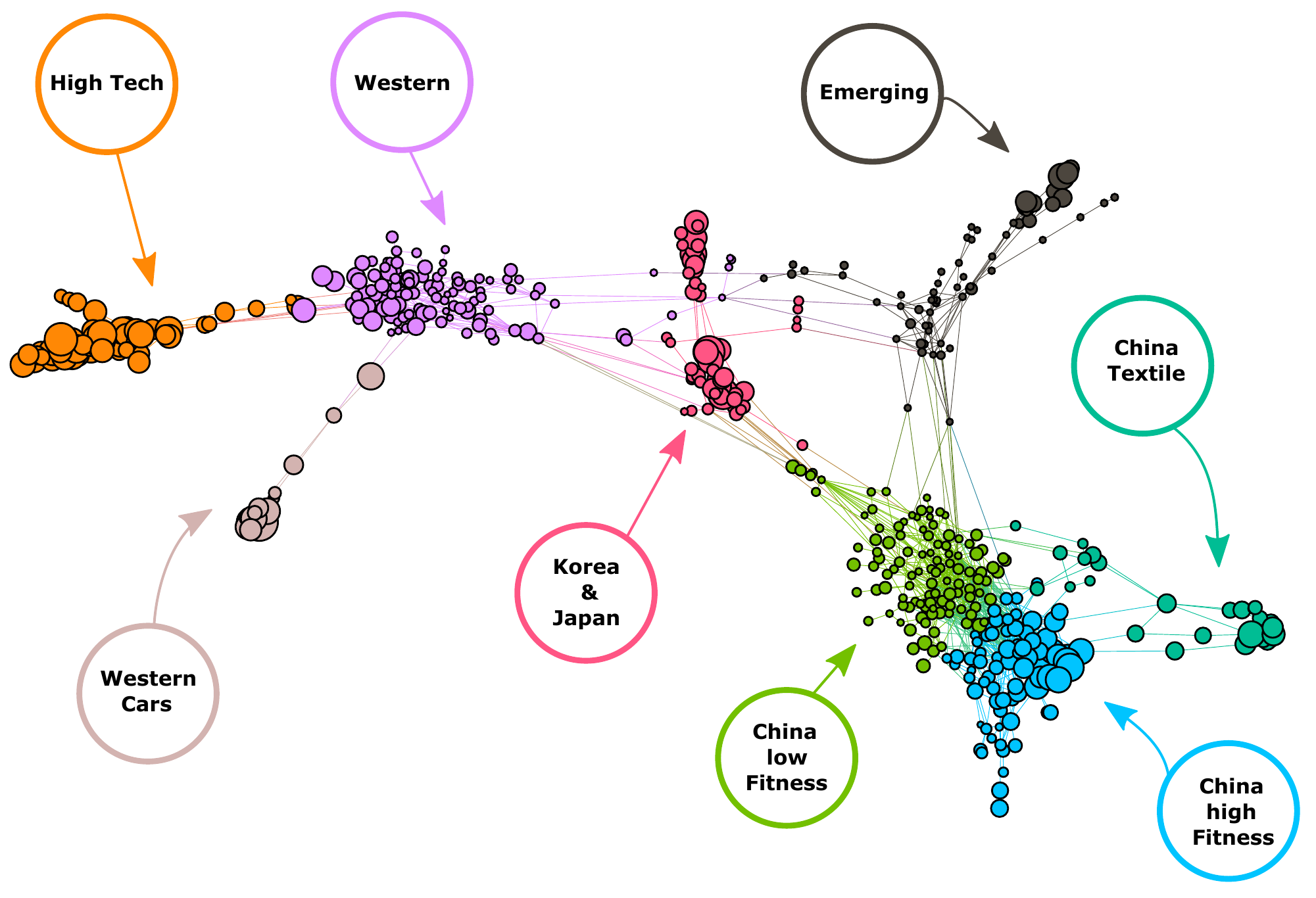}
\caption{\textbf{Projected network of metropolitan areas.} Clusters of MAs obtained through the community detection algorithm: the values of modularity and density are $0.68$ and $0.012$, respectively. Node size is proportional to the fitness of the MA. We notice that the high-tech cluster is the one containing the MAs with the highest fitnesses.}
\label{fig:netw_AM}
\end{figure}
The resulting networks are shown in Fig~\ref{fig:netw_tec} and Fig.~\ref{fig:netw_AM}.\\ 
The technology network of Fig.~\ref{fig:netw_tec} does not show a strong modular structure due to the ability of MAs to produce patents in different areas. Instead, different specific communities, with contiguous clusters containing products of similar macro-type. 
For instance, we can find the technology communities of (clockwise, starting from the left/light green) communication \& information, weapons, printers, domestic technologies, cars, bicycles, buildings, textile, plastic, metallurgy, agri-food \& mining, fuel, organic chemistry, train, nuclear energy and clock. Node sizes are proportional to their complexity. The cluster with the highest number of complex nodes is the communication \& information one, pointing out that not all MAs have the necessary capabilities to patent in this area.\\
Increasing the resolution parameter in the modularity optimisation~\cite{lambiotte2008laplacian}, we can identify three technological macro areas. These three areas correspond respectively to the light blue, pink and olive green nodes in Fig.~\ref{fig:netw_tec} \textbf{(b)}: 
The three regions contain different kinds of technologies:
\begin{itemize}
    \item Car technologies: this region, coloured in olive green, contains technologies closely related to cars;
    \item Highly sophisticated technologies: this macro area, depicted in light blue, contains clusters such as electricity and communications, nuclear, and household items, all technology sectors that we can classify as highly sophisticated technology sectors;
    \item Manufacturing technologies: in this area, represented in pink, we can find clusters related to the textile, agri-food, plastic and paper industries, thus containing manufacturing technology sectors.
\end{itemize}
Finally, in Fig.~\ref{fig:netw_tec}, we colour the technologies to show the average RCA values of New York (\textbf{c}) and Shanghai (\textbf{d}) in the database years. Red implies a higher RCA value, and we can note how Shanghai has focused more on manufacturing technologies while New York is strong in electricity and communications technologies.\\
As for the projected network of MAs (Fig.~\ref{fig:netw_AM}), $\alpha = 0.1$ was used as the significance threshold parameter for the projection of each window; the depicted communities were found via the Louvain algorithm, and the partition features a modularity of $0.68$.
This statistically validated projection shows how the MAs can be grouped according to specific criteria. We find well-defined communities of Chinese, emerging, Euro + USA, Japanese + Korean MAs. We also found a high fitness cluster containing MAs such as London, New York or San Jose, and the Western MAs cars manufacturing including Turin, Detroit and Stuttgart, among others. We can see how the Japanese \& Korean cluster shares connections with MAs from western countries and the Chinese and emerging countries. We present a complete table of MAs with their class in the Supplementary Information.\\

\subsection*{The Fitness-GDP relation in metropolitan areas}
\begin{figure}[h!]
\centering
\subfloat[]
   {\includegraphics[width=.50\textwidth]{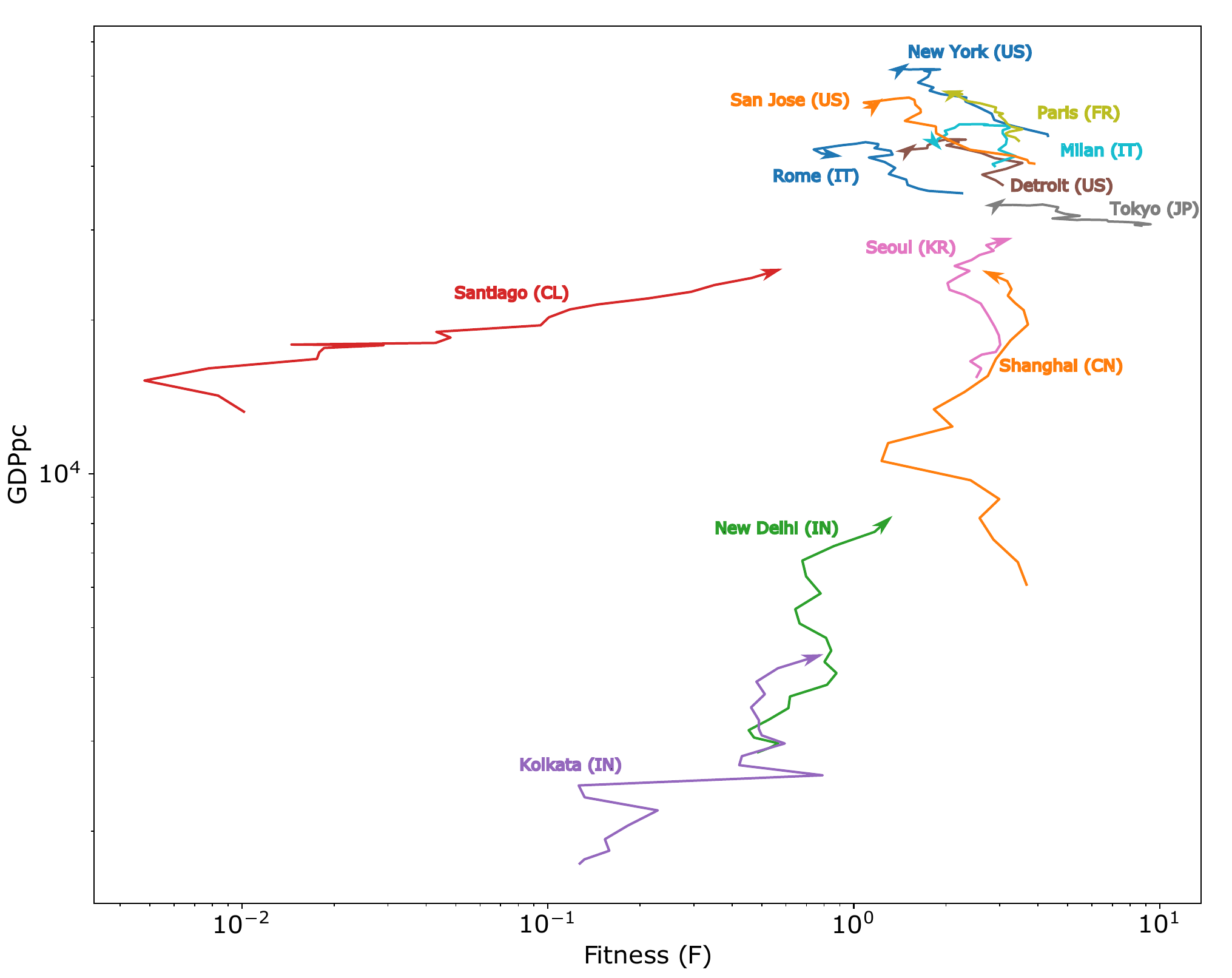}} \quad
\subfloat[]
   {\includegraphics[width=.415\textwidth]{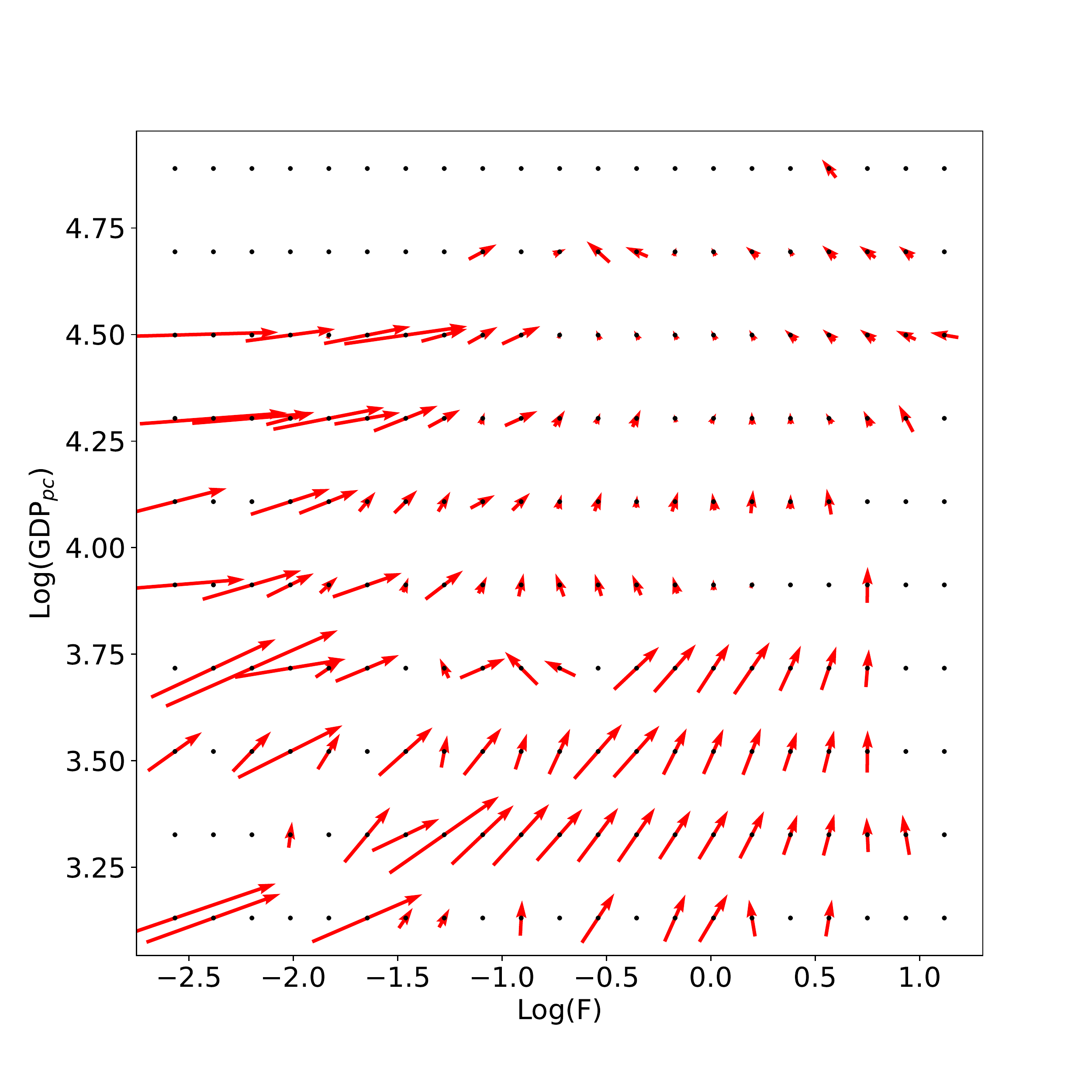}} \quad
\subfloat[]
   {\includegraphics[width=.50\textwidth]{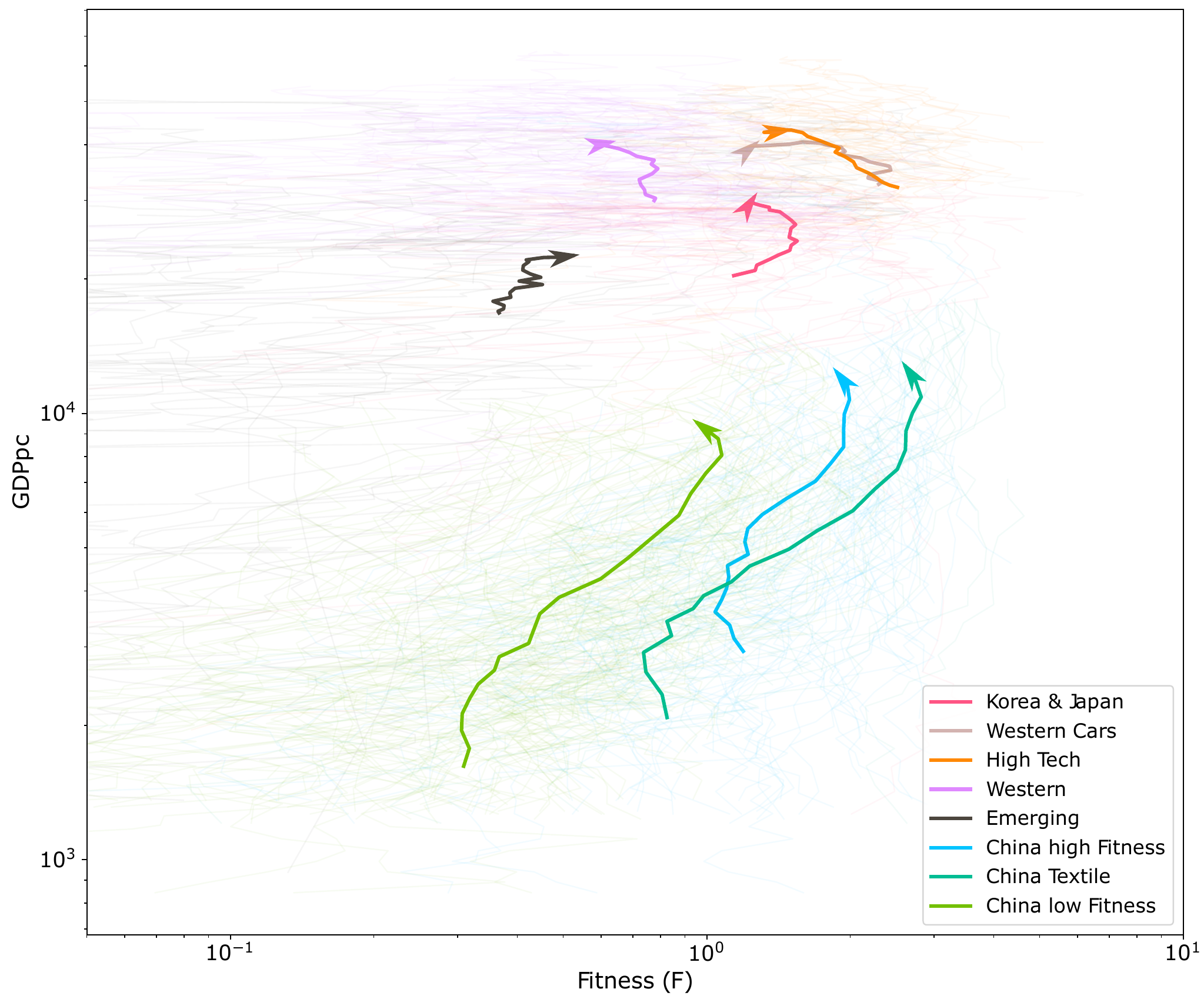}} \quad
\caption{\textbf{The Fitness-GDPpc plane in the case of metropolitan areas and their technological production.} \textbf{(a)}. We trace the trajectory of some MAs from 1990 to 2010 in the Fitness-GDPpc plane. MAs with high fitness show a more significant increase in GDPpc. (\textbf{b}). We show an average vector field of the trajectories from 1995 to 2005. In this plot, we can better visualise how high Fitness leads to increases in GDPpc (most evident in the lower right). In contrast, MAs with low Fitness will tend to increase this first. \textbf{(c)}. Trends of all MAs, with trajectories coloured according to the community of belonging. For each community, we highlighted the average trajectory.}
\label{fig:GDPpc-Fpc}
\end{figure}

In Fig.~\ref{fig:GDPpc-Fpc}, we report the results regarding the relationship between the technology basket of MAs and their GDPpc.\\
We apply the Fitness and Complexity algorithm described in the Methods section: we first calculate the complexity of technologies at the country level and then compute the exogenous Fitness of the MAs.
In Fig.~\ref{fig:GDPpc-Fpc} we report three different representations of the GDPpc-Fitness plane. In the first (\textbf{a}), we trace the trajectories of some MAs from 1990 to 2010. MAs with high fitness are generally more likely to have a more significant increase in GDPpc. For  Shanghai, for instance, the trajectory is nearly vertical, arguably thanks to the high starting fitness, ending at a similar value of GDPpc as Santiago. Santiago is also an interesting case as its trajectory moves in an almost horizontal line increasing the fitness but cannot improve the GDPpc quite as much as Shanghai, arguably due to the low initial fitness. Other MAs, such as the Indian New Delhi and Kolkata, also tend to grow consistently in fitness and GDPpc. 
The same phenomenology is mirrored in Fig.~\ref{fig:GDPpc-Fpc} (\textbf{b}) where we show the average vector field of the trajectories from 1995 to 2005. Here, we also observe that the higher the fitness, the higher the increase in GDPpc. Finally, in Fig.~\ref{fig:GDPpc-Fpc}\textbf{c} we show the overall trend of all MAs whose trajectories are coloured according to the community of belonging. For each community, we highlight the average trajectories. The three communities of Chinese MAs are particularly interesting since they show similar trends of fitness increase. The other clusters show different trends that can be easily interpreted in terms of GDPpc and Fitness. The {\em High-Tech} cluster has the highest average GDPpc, while the {\em Western} and {\em Western cars} ones have the same average GDPpc with the difference that the latter has higher fitness. The cluster Korea \& Japan has a low average GDPpc compared to the previous three clusters, though with a comparable fitness. The cluster labelled as Emerging is slowly increasing in terms of average GDPpc and fitness. Finally, we note that the fitness trends of all clusters are decreasing, except for the Emerging and Chinese clusters. This behaviour is justified by considering that the fitness is a  globally computed quantity, using data about all MAs. For this reason, the fitness cannot increase for all MAs simultaneously, and if it increases for some MAs, it must automatically decrease for others.

\subsection*{The innovation fitness rankings of metropolitan areas}
\begin{figure}[h!]
\centering
\subfloat[]
   {\includegraphics[width=.49\textwidth]{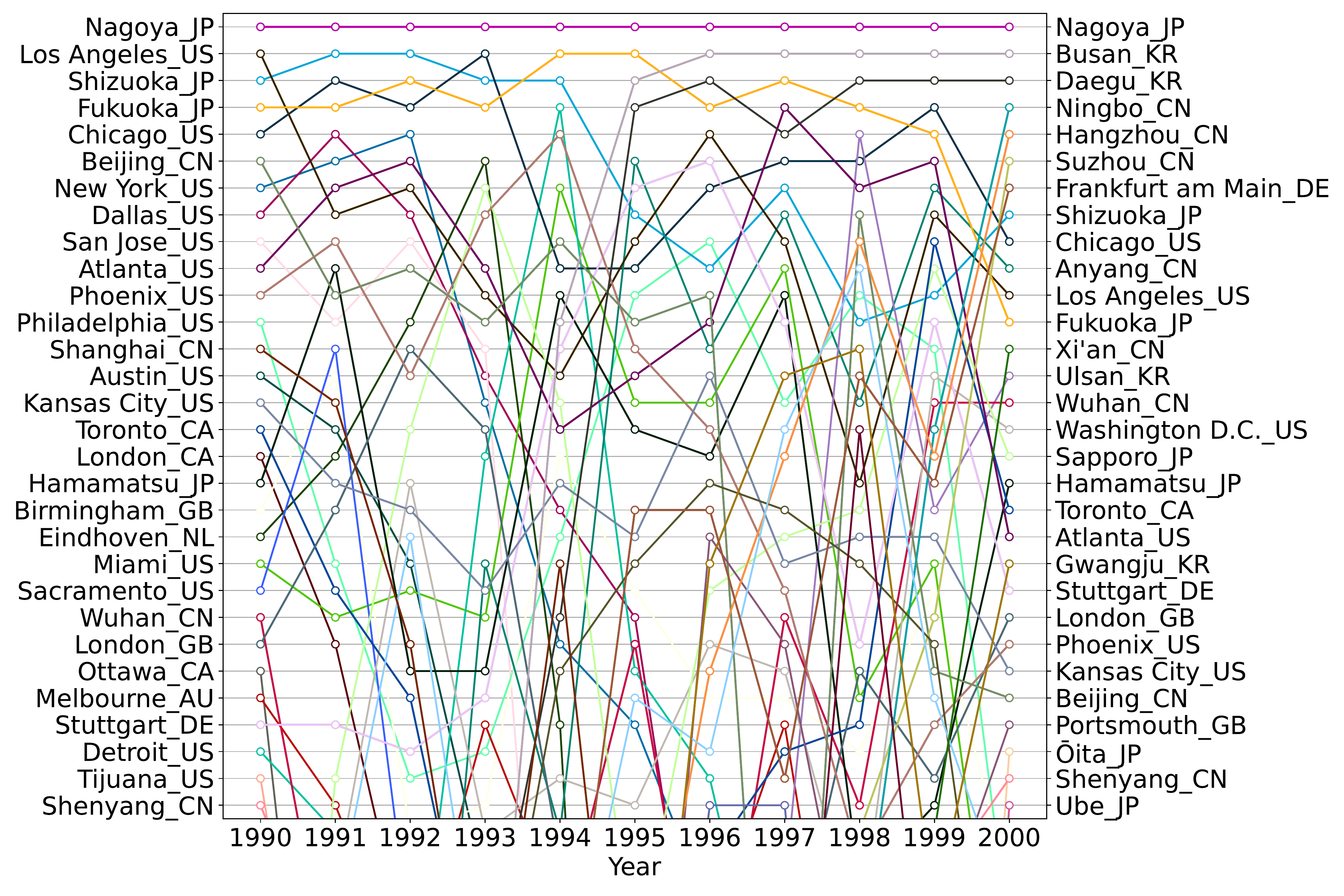}} \quad
\subfloat[]
   {\includegraphics[width=.49\textwidth]{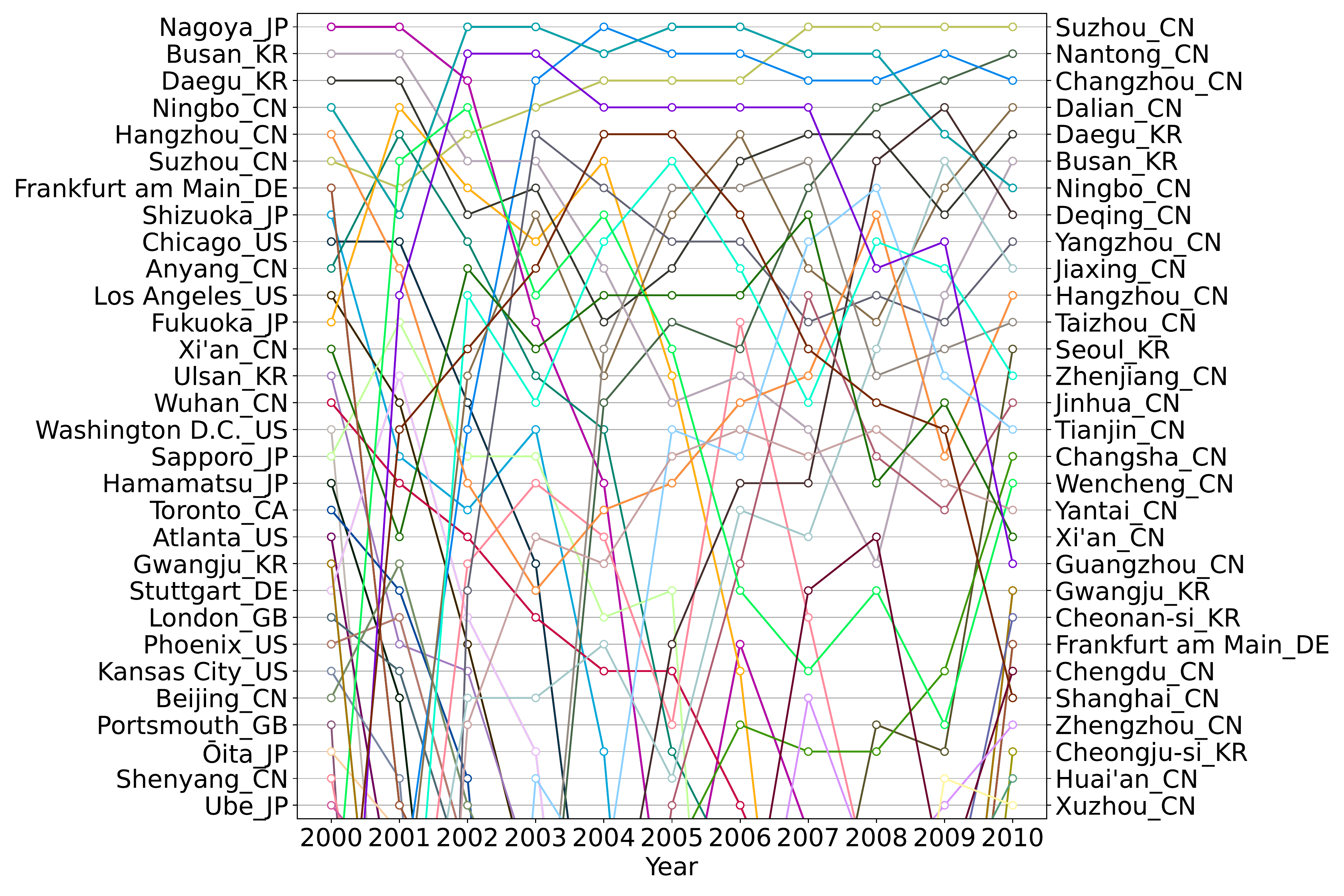}} \quad
\caption{\textbf{The Fitness rankings of metropolitan areas.} The 30 MAs with the highest fitness are shown, along with the evolution from 1990 to 2000 \textbf{(a)} and from 2000 to 2010 \textbf{(b)}. In 1990 many of the metropolitan areas in the top 30 of the fitness rankings were from the US, Europe, Canada and Japan, with only Shanghai and Beijing from China. In 2000, Chinese and Korean MAs appear in the top 30, and in 2010 they dominate the top of the fitness rankings with Frankfurt as the only European, 6 Korean MAs and all others being Chinese.}
\label{fig:ranking}
\end{figure}

The metropolitan areas with the highest fitness per year are presented in  Fig.~\ref{fig:ranking}. It is remarkable, even in this case, the rise of Chinese MAs from 1990 to 2010: at first, only the biggest areas such as Beijing and Shanghai enter in the top 30 of the fitness rankings. Nagoya (Japan) sits atop of the rankings from 1990 to 2001, then it is overtaken by the wave of Chinese cities who start to monopolise the top 30 shortly after 2000; in 2000 the rankings are still mixed, including many Chinese metropolitan areas but also still many from the US and Japan. Ten years later, there are only seven metropolitan areas in the top 30 which are not Chinese: six of these are Korean and only one is European, Frankfurt. In 2020 Suzhou tops the rankings, followed by other Chinese metropolises such as Nantong, and the first non-Chinese MAs are the Korean Daegu and Busan, which were also at the top in the 2000 rankings.

\subsection*{Coherent technology production}
\begin{figure}[h!]
\centering
\subfloat[]
   {\includegraphics[width=.50\textwidth]{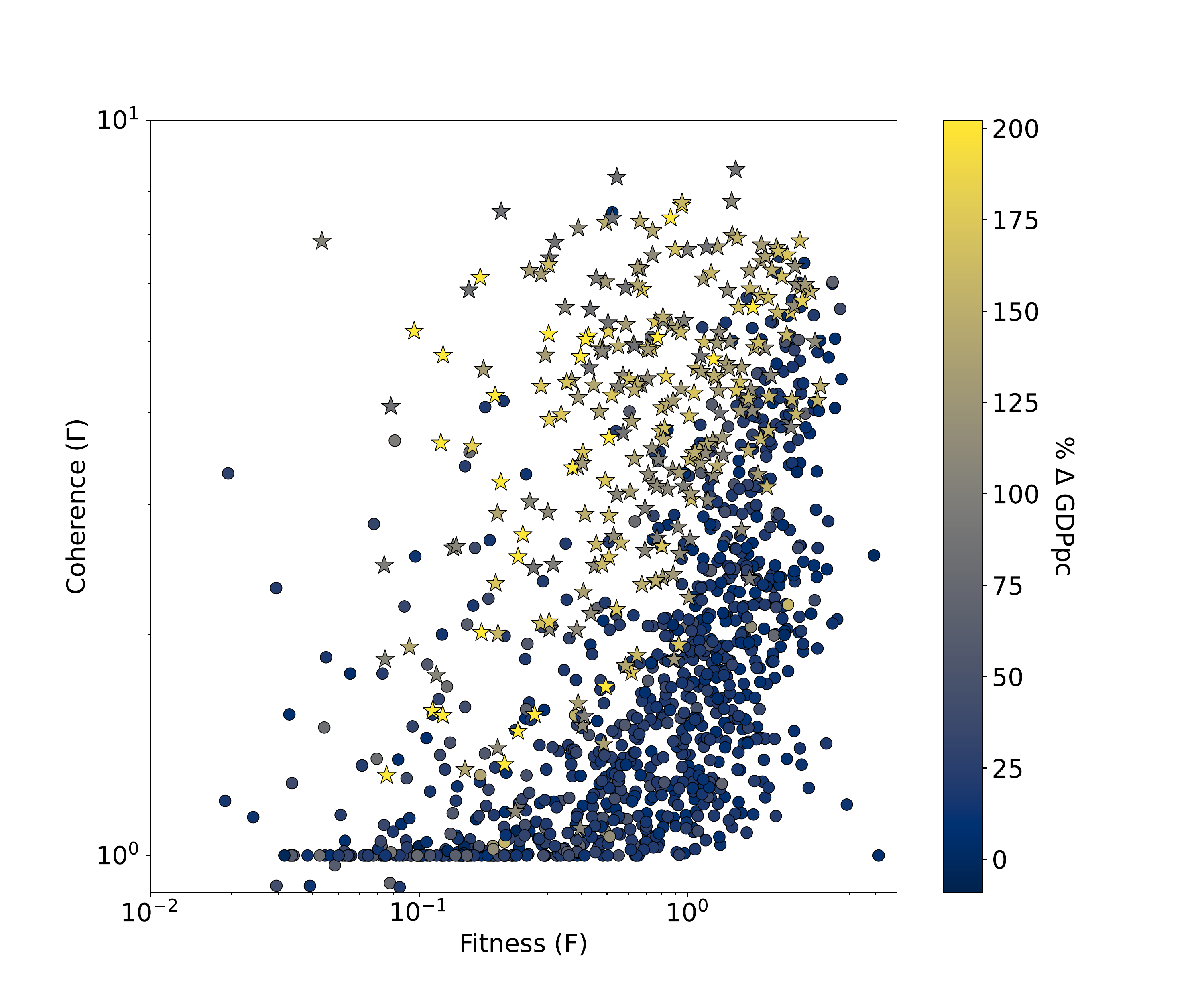}} \quad
\subfloat[]
   {\includegraphics[width=.48\textwidth]{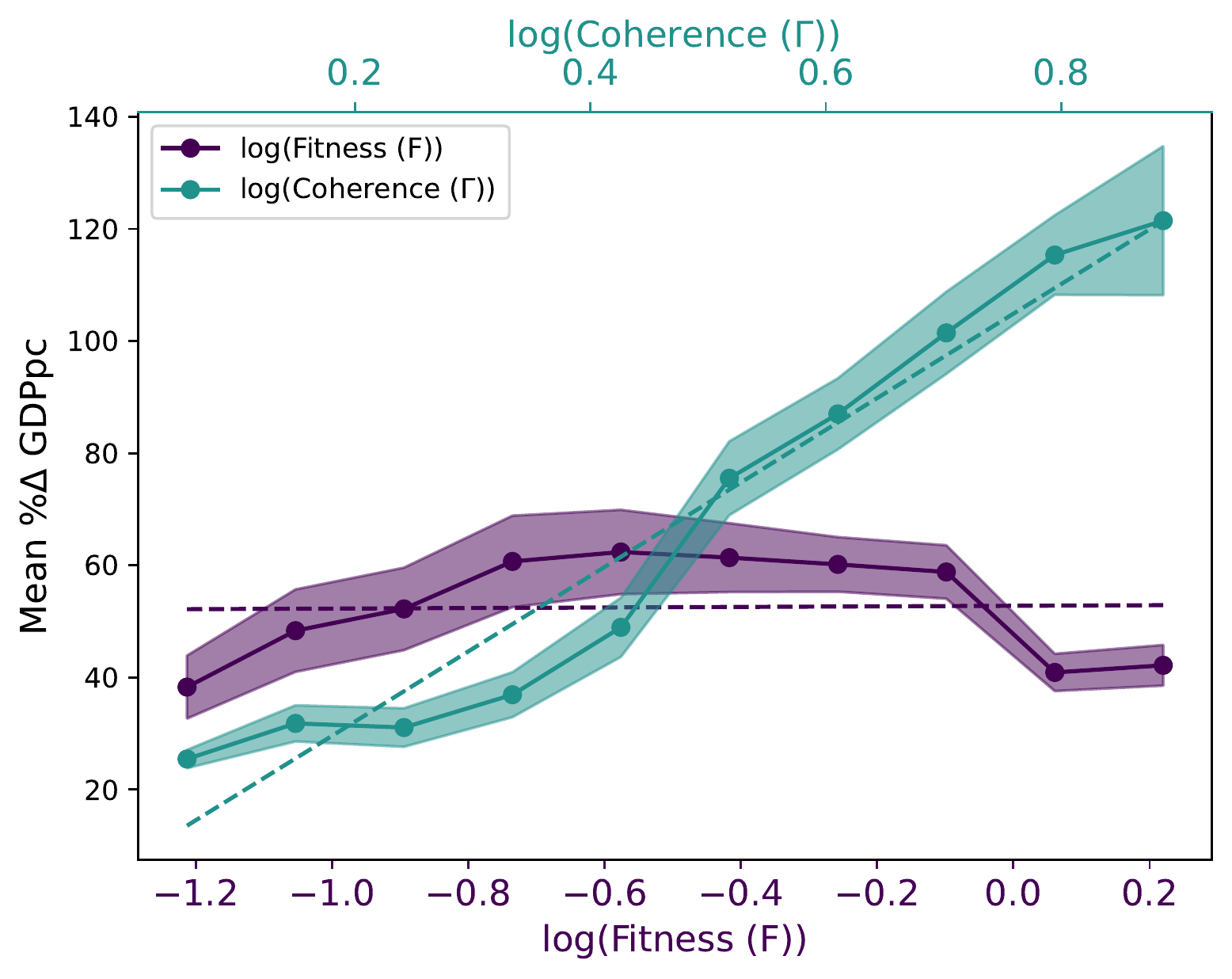}} \quad
\caption{\textbf{Fitness VS Coherence to evaluate GDPpc growth.} \textbf{(a)} Fitness - Coherence plane. We represent the averages of the measures over the decade 1995-2005, and the colour scale is the fractional change of GDPpc over the years.
We observe how the coherence allows discriminating MAs with a more significant positive change in GDPpc. Stars indicate the Chinese MAs. 
\textbf{(b)} Average fractional change of the GDPpc versus Fitness and Coherence. To highlight that coherence can better discriminate changes in GDPpc, we divide fitness and coherence into ten bins and calculate the mean fractional GDPpc variation of all the points in each of the ten bins. We show how fitness and coherence display different behaviours. In particular, the fitness curve is roughly constant, highlighting that the fitness cannot discriminate different fractional changes of the GPDpc. Coherence, instead, displays a growing trend with the fractional change of GDPpc, i.e. the higher the fitness, the higher \% $\Delta$ GDPpc.}
\label{fig:coer}
\end{figure}

In Fig.~\ref{fig:coer} we show the results of the coherent diversification in technological production.
From Fig.~\ref{fig:coer} \textbf{(a)}, displaying the Coherence - Fitness plane, we observe that coherence can capture the signal of significant positive change in the GDPpc of MAs. Fig.~\ref{fig:coer} \textbf{(b)} confirms this picture: while the change in GDPpc is not sensitive to fitness changes, a growing trend of coherence is accompanied by a parallel growth in the GDPpc's change. This result is appealing, especially if we consider that, in the ranking of $\Gamma$, $79$ MAs, out of the top $100$, are Chinese. To ensure that our result is not simply due to the relatively high number of Chinese MAs in our dataset, we performed a robustness test described in more detail in the Supplementary Information. In this test, we rebuild the technology network as explained in the "Networks projection" Section without the Chinese MAs, to then compute the Coherence using all MAs.\\ 
In the Supplementary Information, we also ran a simple check to show 
that the high Coherence is not related to low diversification.
The coherent diversification strategy of China was already highlighted in a previous work by Gao et al.~\cite{gao2021spillovers}, who noticed similar coherent patterns for the expansion of the production in Chinese regions.

\section{Discussion}\label{sec:discussion}
In this work, we studied technological innovation in metropolitan areas by analysing data on the production of patents. In particular, we focused on the signals of specialisation and diversification by applying the Fitness and Complexity framework and novel methods for bipartite networks to the technological production of metropolitan areas. The Fitness and Complexity algorithm application for MAs is particularly interesting since the interplay between specialisation and diversification can change at different scales~\cite{laudati2022different}.\\
We found that MAs tend to specialise in technology sectors, particularly for some technological categories, such as cars or electronics. 
Moreover, we observed similarities among metropolitan areas within a country or across similar countries. Chinese MAs give the best example of similar MAs in a single country. They are organised in three coherent clusters specialised in similar technological baskets. One of the clusters is specialising in the technology sectors of textile industries, 
another one specialises in agri-food 
and the third cluster is devoted to highly sophisticated technology sectors.
We observe a similar behaviour of relatedness, though at a smaller scale, in Japanese and South-Korean MAs. We also observe similarities among emerging MAs and among highly technological metropolitan areas. Interestingly, the network of similarities among MAs shows a clear geographical boundary between highly developed Asian and Western (European/American) MAs.\\
We used the Fitness and Complexity framework to understand the economic evolution of MAs and their clusters.  In line with previous results, we have shown that Fitness can drive an increase in GDP per capita: MAs with a complex technological basket tend to have higher GDPpc in the following years than MAs developing more basic technologies. Korea and Japan followed this path, especially in the past. In recent years the standout case is China: the complexity of innovation in Chinese MAs is very high, and their GDPpc displays rapid growth. We found that Chinese metropolitan areas are not only able to diversify their innovation patterns by aiming for a more complex technological basket but also do this in a coherent and coordinated way. Measuring, in fact, the coherence of the innovation baskets of MAs, we show that a vast majority of MAs with the highest coherence values are Chinese, and we report that this outcome is not due to a restriction to a specific set of technologies. On the contrary, Chinese MAs diversify consistently and coherently. Moreover, a coordinated effort is also evident, with Chinese MAs areas sharing common sets of technologies. We found that coherent diversification is necessary and arguably as important as fitness to increase the wealth of a metropolitan area, as the highest increase in GDPpc is found in metropolitan areas with both high fitness and coherence.\\
We found that from 1990 to 2010, the top 30 items of the patents production's Fitness rankings drastically changed. In 1990, many MAs from many rich countries were sitting at the top of the table, with Japan and the US vastly represented, Nagoya and Los Angeles in the top two positions, and only Beijing and Shanghai as Chinese metropolitan areas. By contrast, in 2010, only seven metropolitan areas in the fitness top 30 rankings were not Chinese: six Korean ones and a European, Frankfurt, with the Chinese Suzhou topping the table.

The theoretical framework presented here can be applied in several scenarios for investigating questions arising from our analysis:

\textit{Optimal diversification strategies and technology forecasting for MAs at different scales and capabilities.} Our theoretical framework can be applied to study the best diversification strategy for MAs, assessing the best technologies to develop in a city, as in~\cite{straccamore2022will,albora2021product}. However, some metropolitan areas can diversify as much as they want because their size and capabilities are close to those of a whole country; others cannot diversify their technological products as much because they do not have the resources to do so. \textit{Specialisation} and \textit{diversification} are both feasible ways for MAs to compete depending on their resources, acting more as a large firm~\cite{laudati2022different} or as a whole country~\cite{tacchella2012new}.

\textit{The strategy of Chinese MAs.} We found that the Chinese MAs have the most \textit{coherent} technology diversification and specialisation strategies. These results align with other work such as~\cite{long2012patterns}, but the cause of the observed structured diversification remains unanswered: is this behaviour coordinated nationally? A more detailed analysis of the Chinese case could highlight whether China is implementing a long-term, all-purpose strategy for developing technologies, defining a priori the production basket of single MAs. If this is true, can the strategy be copied by other countries, and under which conditions? For instance, some emerging MAs, such as Indian ones, were on a trajectory similar to Chinese MAs, as shown in Fig.~\ref{fig:netw_AM}\textbf{(a)} for New Delhi or Kolkata. 

\textit{The restricted business of car technologies}. The strong signal from MAs dedicated to producing cars is unique and suggests that these metropolitan areas could have trouble diversifying their production. It is not clear yet whether this is a signal of high competitiveness of these kinds of technologies, and therefore MAs should specialise to better profit from this production, or it is hard to implement other technologies for car-focused MAs. However, with the advent of electric cars and considering the significant technological changes about to occur in the forthcoming years (see, for instance, the European ban on fossil-fuel car production by the EU\footnote{\url{https://www.euronews.com/green/2022/05/12/eu-wide-ban-on-new-fossil-fuel-cars-to-kick-in-from-2035-as-lawmakers-back-proposal}}), the future economy of MAs currently producing cars will have to be reshaped. Future studies focusing on optimal diversification strategies and forecasting future technology production could be used to shape technology paths that can help these MAs adapt to such changes.



\bibliography{sample.bib}

\section{Data availability statement}
The data that support the findings of this study are available upon reasonable request from the authors.

\section{Acknowledgements}
We thank Angelica Sbardella for insightful and useful discussions. All the authors acknowledge the CREF project ‘Complessità in Economia’.

\end{document}


\title{\huge{Supplementary Information}\\\Large Urban Economic Fitness and Complexity from Patent Data}

\author{Matteo Straccamore, Matteo Bruno, Bernardo Monechi\\ and Vittorio Loreto}

\date{}
\maketitle

\newpage

\tableofcontents
\newpage

\newpage

\section{MA data cleaning and statistical data visualization.}
In the main text we described how to obtain the final matrices composed of $2169$ MAs and $650$ technology codes. However, to make the data cleaner, we decided to cut out from the analysis those MAs that are not very active in technology production. We decided to quantify their activity by setting a threshold for MAs as follows:
\begin{equation*}
    \overline{RCA}_a = \frac{
    \sum_{y\in Y}\sum_t{RCA^y_{at}}}{|Y|}
    < 1.
\end{equation*}
We denote by $\textbf{RCA}^y$ the RCA matrix obtained by calculating the RCA values from the $\textbf{V}^y$ matrix defined in the main text. The matrix element $RCA_{at}^y$ denotes the value of RCA related to the technology $t$ of MA $a$ developed in year $y$. By $Y$ we denote the set of years in the database.
We then denote by $\overline{RCA}_a$ the average over the years of the overall RCA of MA $a$, i.e., the average over the years of the RCA obtained by summing the RCA values of the entire technology portfolio of $a$. Put simply, we filter MAs that on average are relevant innovators in less than one technology per year.
\begin{figure}[h!]
    \centering
    \includegraphics[scale=.5]{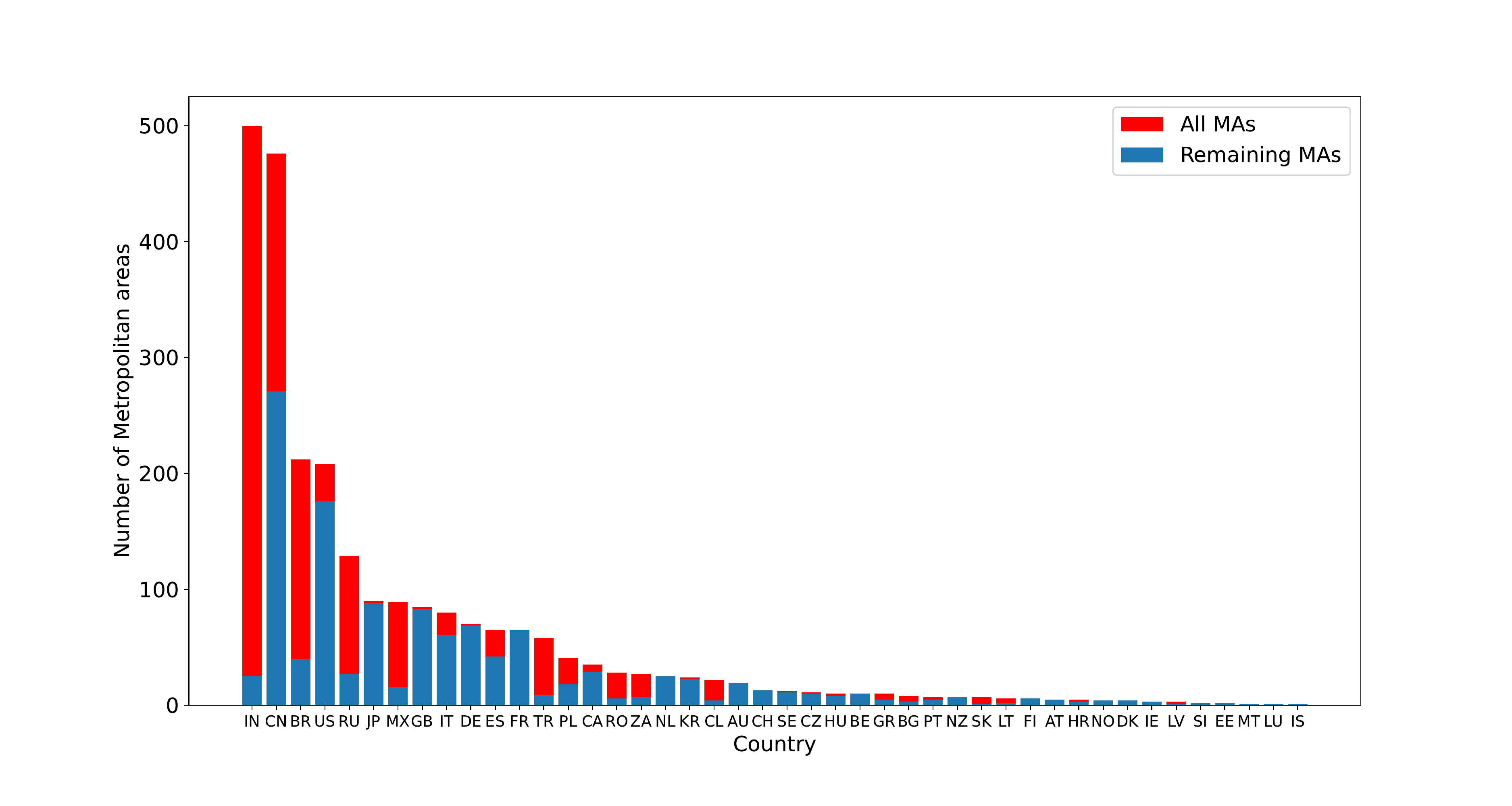}
\caption{\textbf{Distribution of the number of MAs for each country in the database.} With red colour we indicate the MAs in the database that are cut off, in blue those that remain. We use as a cut-off criterion the fact that if the average total RCA across all years of an MA is greater than $1$ then the MA remains. We want to emphasise that we are interested in capturing relatedness among technologies, and to do this we remove from the analysis the MAs that on average make few technologies.}
\label{fig:distr_c1}
\end{figure}
\\
The remaining metropolitan areas are $1211$ and are distributed as shown in Fig. \ref{fig:distr_c1} and \ref{fig:distr_c2}. We show in these Figures the geographical distribution of MAs available to us around the world \ref{fig:distr_c2} and the distribution of the number of MAs for each country in the database \ref{fig:distr_c1}. In both figures, with red colour we indicate the MAs in the database that are cut off (i.e. all $a$ that have $\overline{RCA}_a < 1$), in blue those that remain (i.e. all $a$ that have $\overline{RCA}_a \ge 1$). We are interested in capturing relatedness among technologies, and to do this we remove from the analysis the MAs that on average make very few technologies. Most of these are from emerging countries, and it will be interesting to study in future work how we can characterise their technological growth. However, in the present work these MAs are not able to help us capture relatedness between technology codes.
\begin{figure}[h!]
    \centering
    \includegraphics[scale=.85]{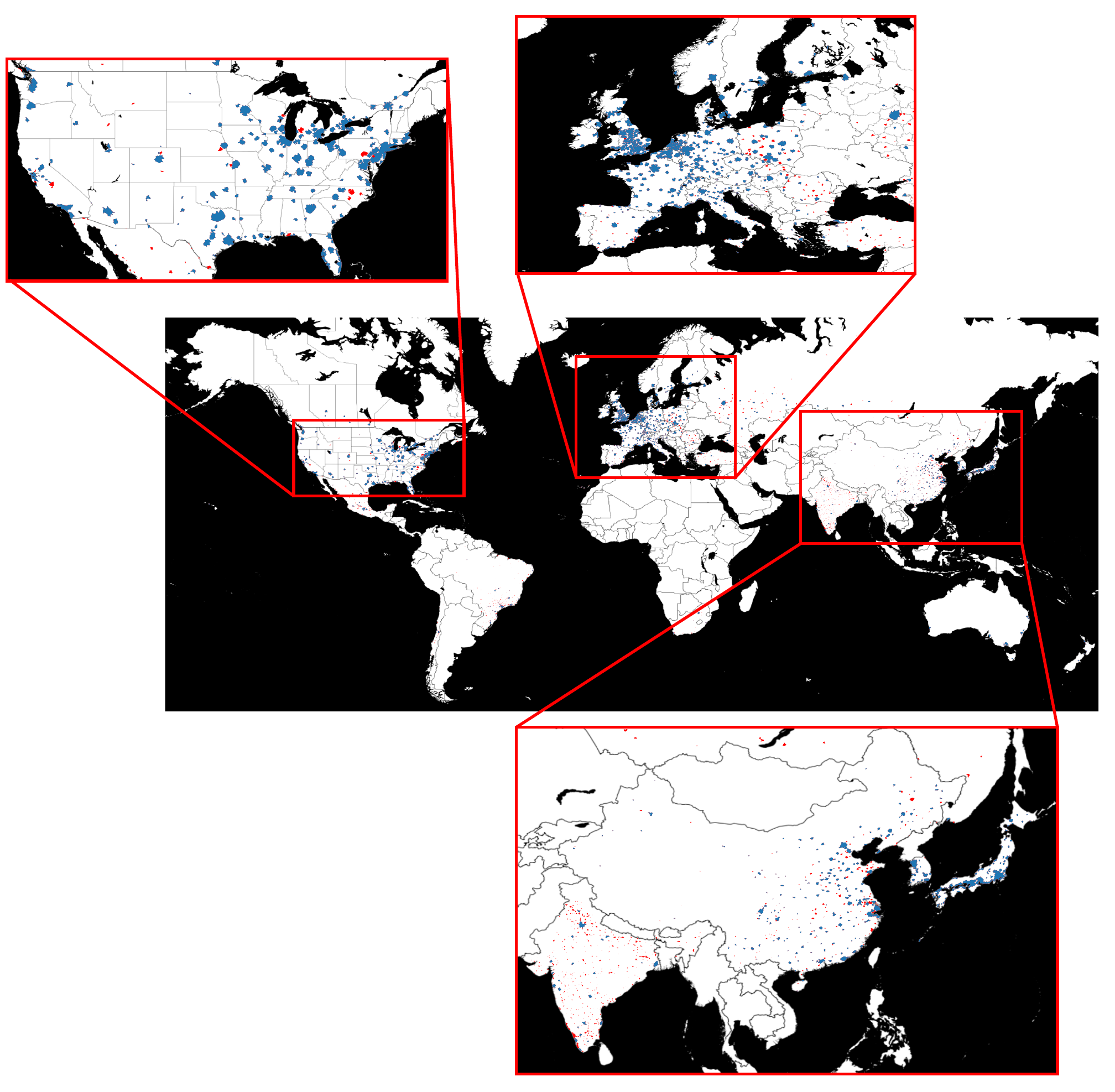}
\caption{\textbf{Geographical distribution of MAs available to us around the world.} With red color we indicate the MAs in the database that are cut off, in blue those that remain. We use as a cut-off criterion the fact that if the average total RCA across all years of an MA is greater than $1$ then the MA remains.}
\label{fig:distr_c2}
\end{figure}

\section{GDP per capita grid example}
In this section we show an example of our GDPpc calculation using the grid constructed in the work of of Kummu et al. [1].
\begin{figure}[h!]
    \centering
    \includegraphics[scale=.35]{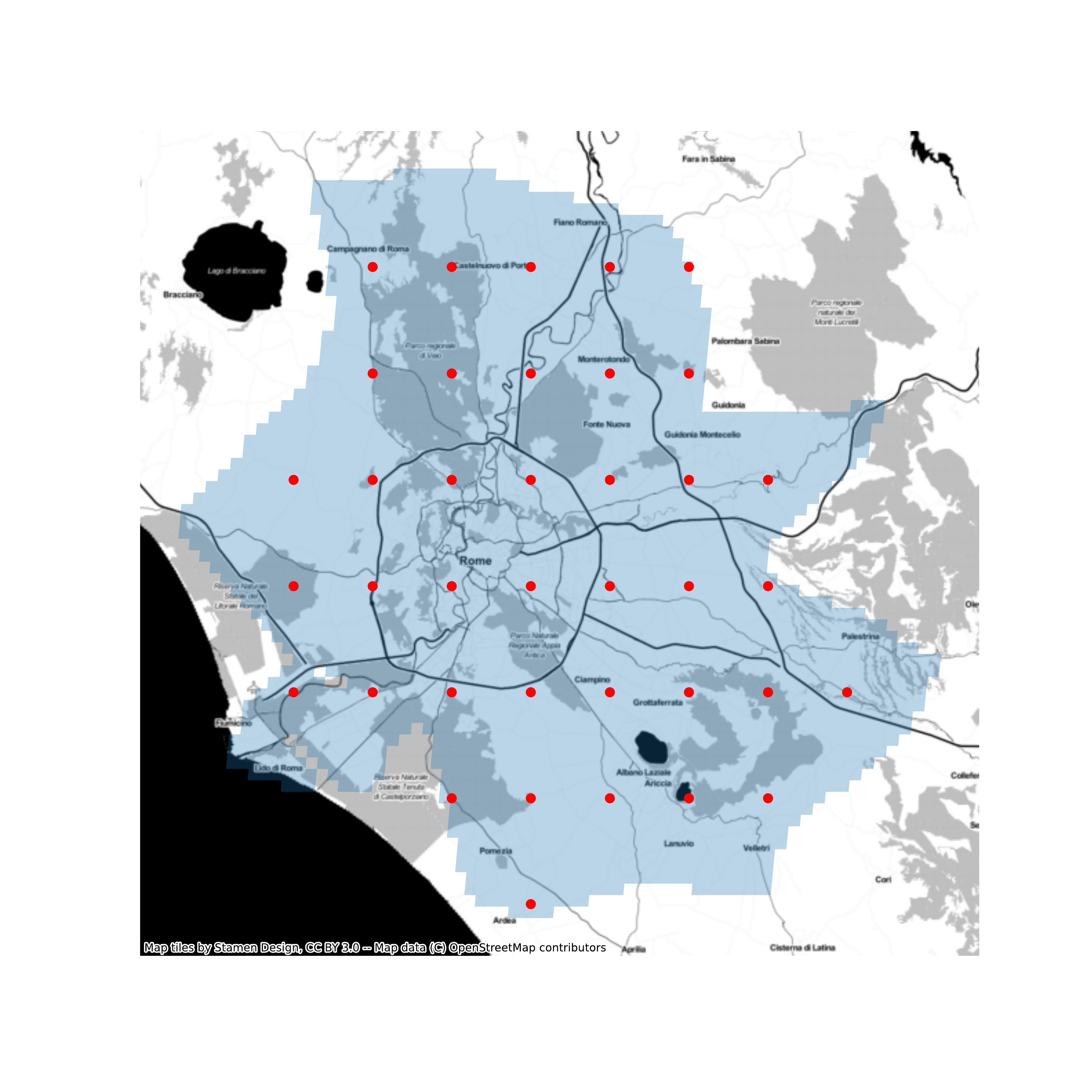}
\caption{\textbf{Example of GDPpc grid in the MA of Rome.} The light blue area is the MA of Rome and the red dots are the part of the grid that intersects the MA.}
\label{fig:grid_ex}
\end{figure}
In Fig.~\ref{fig:grid_ex} we show the intersection of the grid with the MA of Rome as an example. To calculate GDPpc we average all the points (the red dots referring to the figure) within the MA and, because we are working in 5-year windows, we also average over time in each window.

\section{Robustness Coherence test}
\begin{figure}[h!]
    \centering
    \includegraphics[scale=.45]{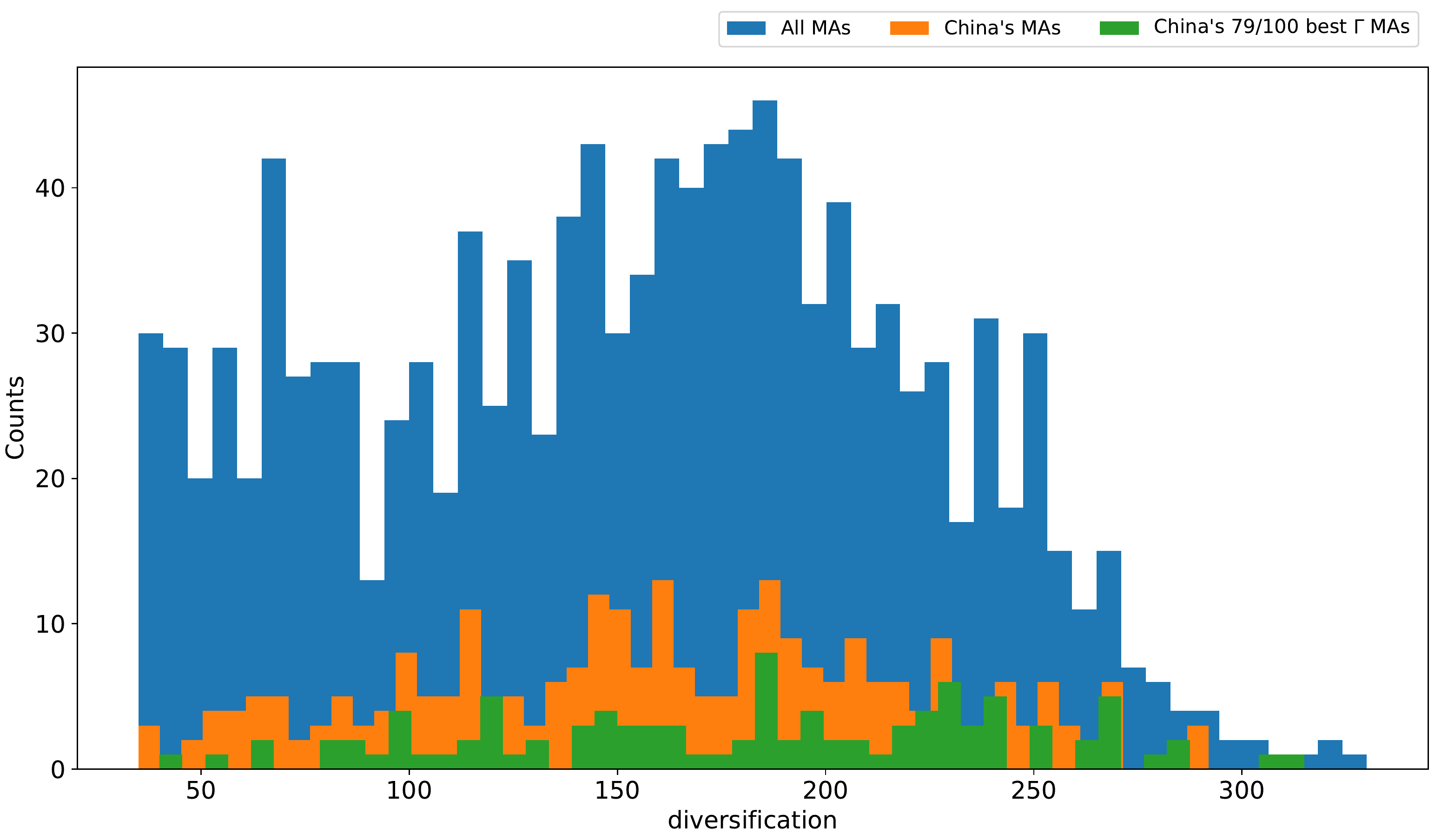}
\caption{\textbf{Diversification distribution.} We plot respectively the distribution of all the MAs, Chinese ones and of the $79$ higher Coherence ones, to show that the high Coherence is not cause to the low diversification.}
\label{fig:d_C_NY}
\end{figure}

In this section we present the coherence robustness test presented in the main text. This consists of showing the diversification distributions of MAs and calculating the technology network without Chinese MAs.\\
In figure \ref{fig:d_C_NY}, we plot the three distributions.
It might happen that Chinese MAs produce few technologies which are closely related; on the other hand, from the trend (green distribution in Figure \ref{fig:d_C_NY}) we see that the 79 Chinese MAs that show a high consistency value are distributed across the diversification spectrum.\\
In the "technology network without Chinese MAs test", we recalculate the network of technologies in the same way that the one in the main text was calculated, except that we do so by removing all Chinese MAs.
\begin{figure}[h!]
    \centering
    \includegraphics[scale=.4]{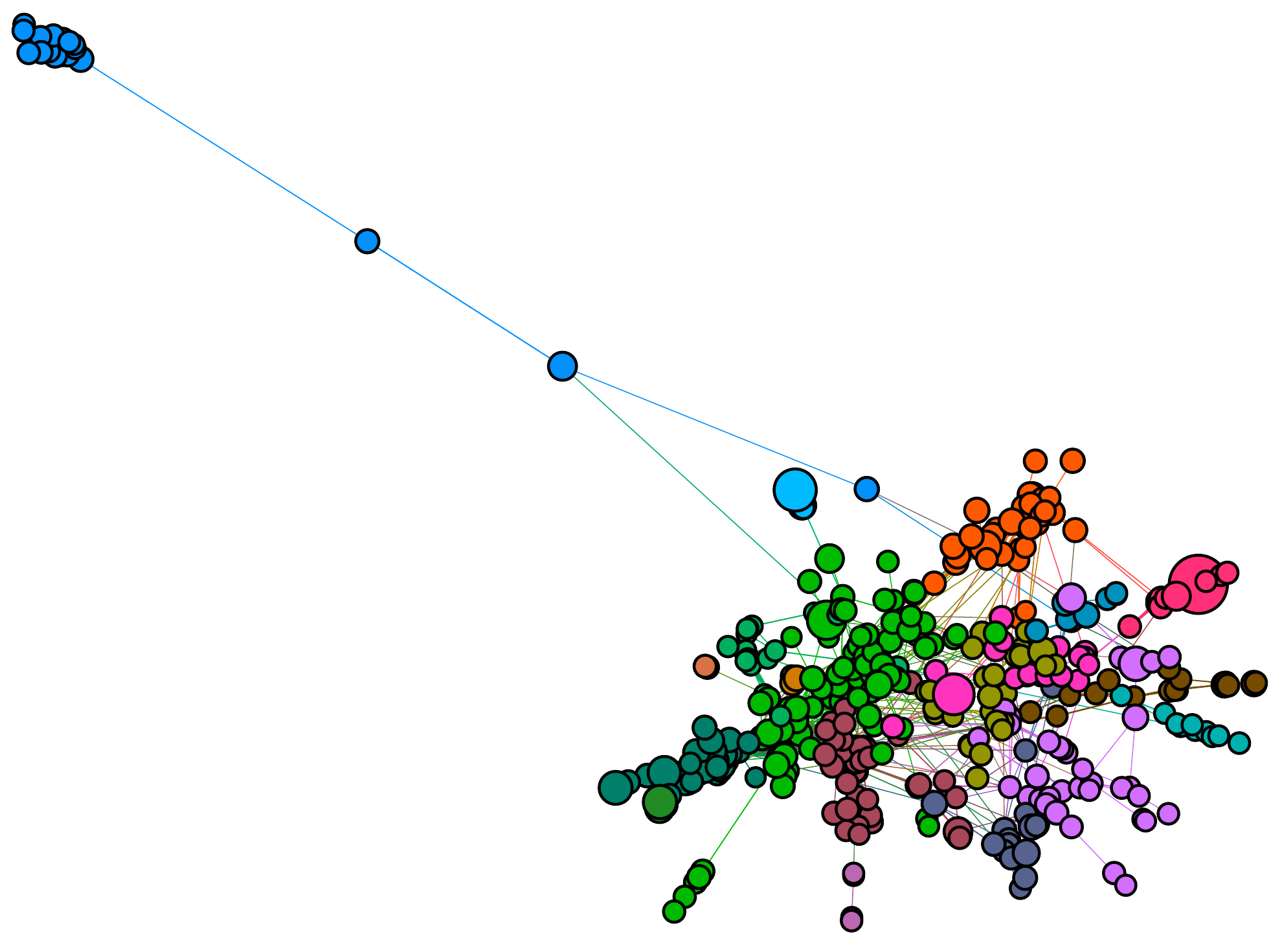}
\caption{\textbf{Technology code network obtained without Chinese MAs.} The size of the nodes is proportional to the complexity of the technology.}
\label{fig:ntNC}
\end{figure}
In Figure \ref{fig:ntNC} we show a representation of the network of technologies obtained without the Chinese MAs.
Starting from this, finally we can compute the coherence diversification of MAs by Equation 6 and 7 in the main text.\\
\begin{figure}[h!]
\centering
\subfloat[]
   {\includegraphics[width=.45\textwidth]{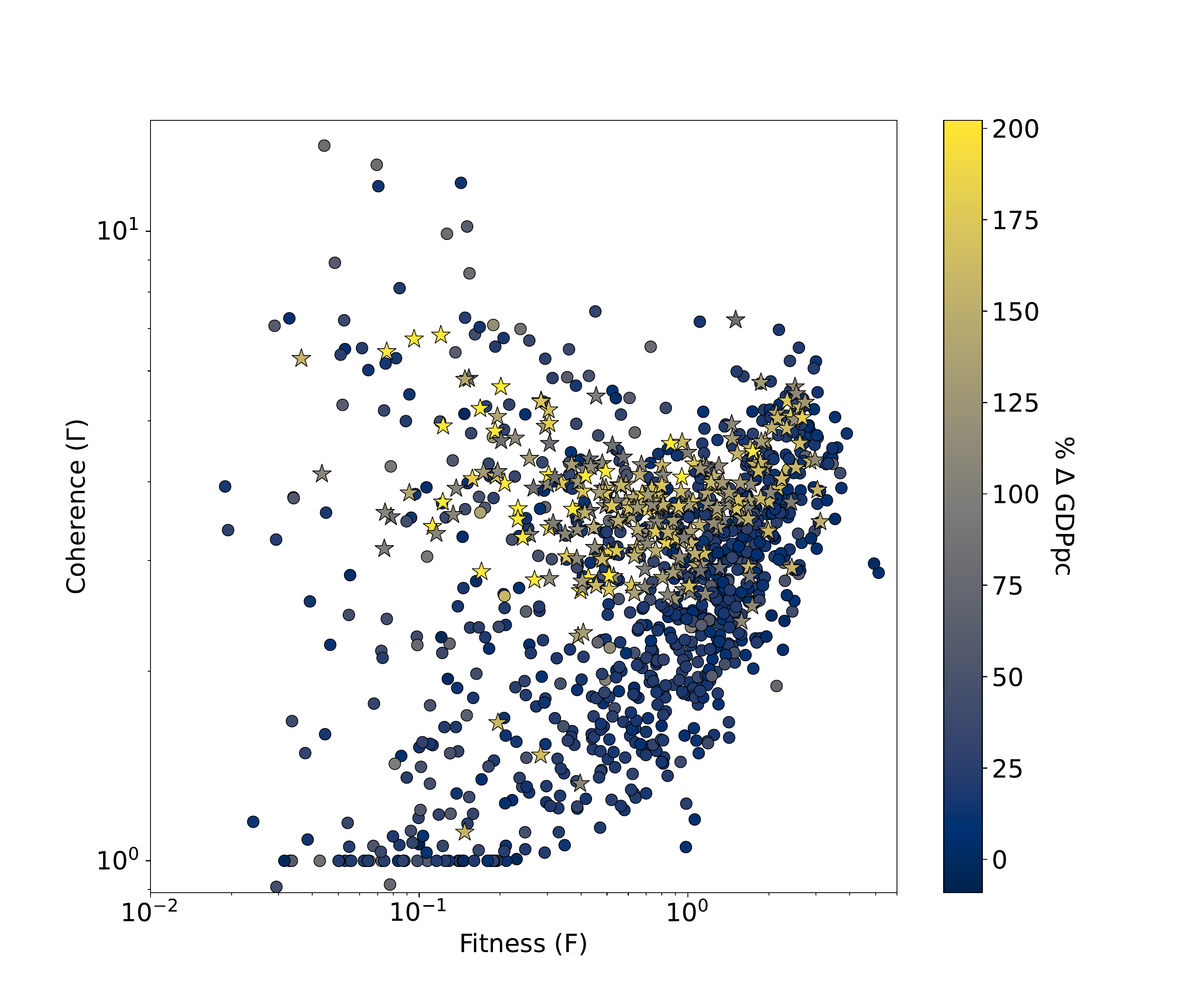}} \quad
\subfloat[]
   {\includegraphics[width=.5\textwidth]{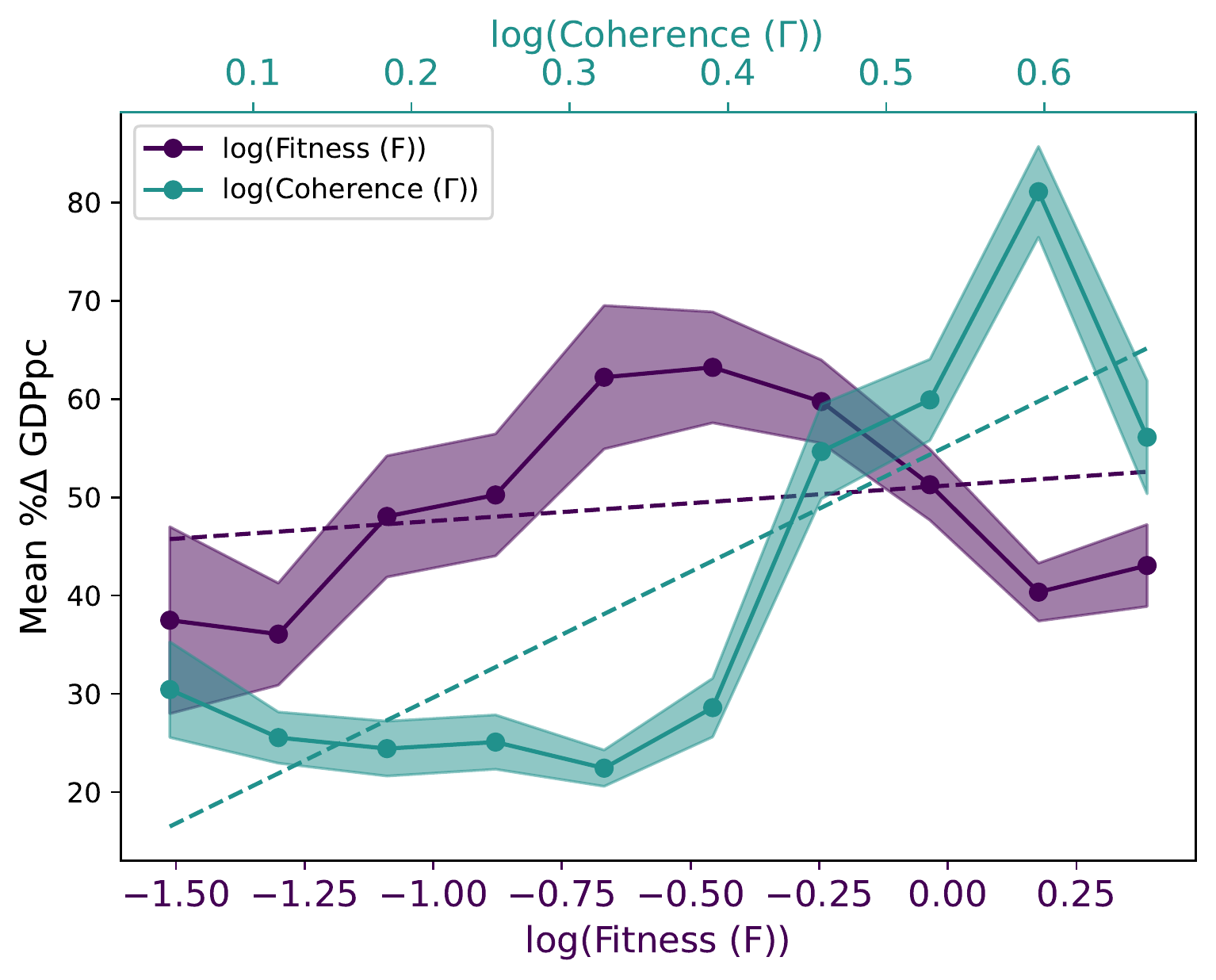}} \quad
\caption{\textbf{Fitness VS Coherence to evaluate GDPpc growth.} \textbf{(a)} Coherence $\Gamma$ VS Fitness $F$ plane. Coherence coefficients (i.e. products similarities) are computed after removing Chinese MAs from the system. Values are the respectively average over the decade 1995-2005 and the color scale is the percentage change of GDPpc over these years. With star markers we are indicating the Chinese MAs. Considering the ranking of $\Gamma$ for MAs, it appears that of the top $100$, $24$ are Chinese ones.
\textbf{(b)} Average GDPpc versus Fitness and Coherence. Again, coherence coefficients are computed after removing Chinese MAs from the system. To highlight that Coherence is better able to discriminate change in GDPpc, we divide both Fitness and Coherence into ten bins and calculate the mean GDPpc variation of all the points in each of the ten bins. We show how the Fitness curve is roughly constant while the Coherence one has an increasing trend. This means that, on average, looking at the Fitness, we will find approximately the same value of \% $\Delta$ GDPpc while looking at the Coherence, the higher this one is, the higher the variation of GDPpc. We show also the lines of best fit of both curves to highlight the difference in the two trends.}
\label{fig:coerSI}
\end{figure}
Finally, in Figure \ref{fig:coerSI} we show the Coherence $\Gamma$ VS Fitness $F$ plane. Values are the respectively average over the decade 1995-2005 and the color scale is the percentage change of GDPpc over this years. We represent Chinese MAs with star markers. Considering the ranking of $\Gamma$ for MAs, it appears that of the top $100$, $24$ are Chinese ones. To evidence that Coherence is better able to discriminate change in GDPpc, we divide both Fitness and Coherence into ten bins and calculate the mean GDPpc variation of all the points in each of the ten bins in Fig. \ref{fig:coerSI}\textbf{b}. We show how the Fitness curve is roughly constant while the Coherence one has an increasing trend. This means that, on average, looking at the Fitness, we will find approximately the same value of \% $\Delta$ GDPpc while looking at the Coherence, the higher this one is, the higher the variation of GDPpc. We show also the lines of best fit of both curves to highlight the difference in the two trends.

\section{Coherence of clusters}

\begin{figure}[h!]
\centering
\subfloat
   {\includegraphics[width=.9\textwidth]{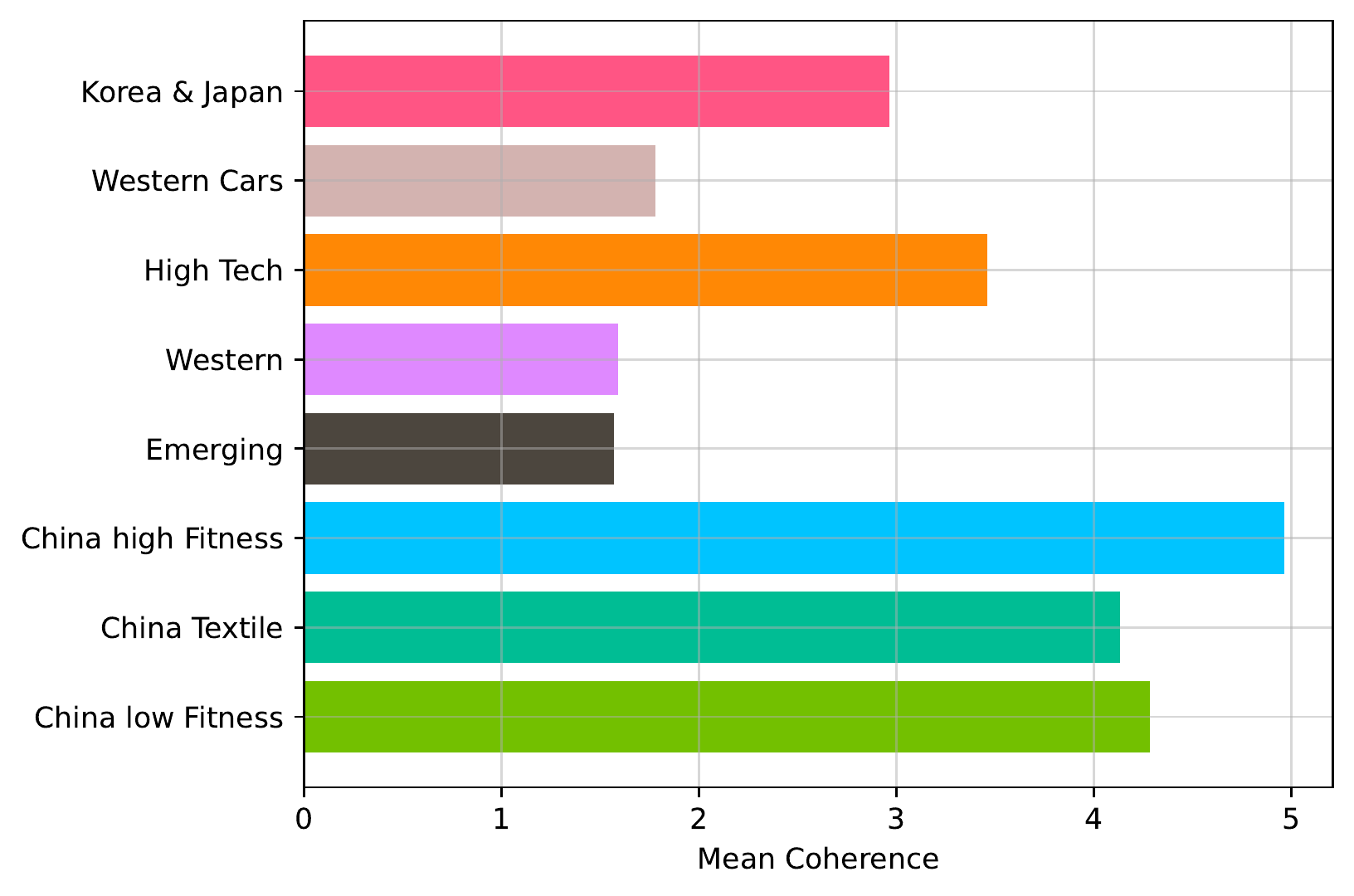}}
\caption{\textbf{The mean coherence of each cluster of metropolitan areas.} The clusters containing Chinese cities have the highest average coherence, and high fitness clusters tend to have higher values of coherence as well.}
\label{fig:mean_coherence}
\end{figure}

In Figure~\ref{fig:mean_coherence}, we show the average coherence for each of the clusters we found, obtained by averaging all coherence values of the MAs in the respective clusters.

Confirming our finding, the Chinese clusters still show the highest levels of coherence. High fitness clusters seem to have higher values of coherence: the high fitness clusters such as high tech or the Japanese/Korean one show on average a higher coherence than the other non-Chinese clusters.

\newpage

\section{Metropolitan areas community class table}
In this section, we show the table in which all the MAs related to the network in Figure 3 of the main paper are present. For each community, we write the country to which the MA belongs.\\

\fontsize{3}{6}\selectfont
\begin{multicols*}{4}
\TrickSupertabularIntoMulticols
\begin{supertabular}[lcl]
{p{0.3cm}p{2.2cm}}
\hline
     \textbf{Country} & \: \: \textbf{Community}\\
\hline
    \: & \textbf{\textit{China Textile}}\\
\hline
    CN & Changzhi \\
       & Changzhou \\
       &   Jieyang \\
       & Guangzhou \\
       &    Heyuan \\
       &   Huaibei \\
       & Tianjia'an \\
       &    Deqing \\
       &  Yangzhou \\
       &   Taizhou \\
       &   Jiaxing \\
       &    Jinhua \\
       &    Jining \\
       &  Quanzhou \\
       &    Suzhou \\
       &   Nantong \\
       &    Ningbo \\
       &    Tai'an \\
       &  Yancheng \\
       & Shuangyashan \\
       &    Xuzhou \\
       &      Jinan\\
       & Zhenjiang \\
       & Zhongshan \\
       &   Yichang \\
       &  Wencheng \\
\hline
\: & \textbf{\textit{China high Fitness}}\\
\hline
    CH & Montreux \\
\cline{1-1}
    CN & Anshan \\
       & Anshun \\
       & Baiyin \\
       &  Baoji \\
       & Baotou \\
       & Beijing \\
       & Binzhou \\
       & Changchun \\
       & Taoyuan \\
       &  Changsha \\
       & Chengdu \\
       & Dalian \\
       &  Ranghulu \\
       & Deyang \\
       & Dezhou \\
       & Dongying\\
       & Fushun \\
       &  Fuxin \\
       & Fuzhou \\
       & Guilin \\
       & Guiyang \\
       & Haikou \\
       & Handan \\
       &  Hangzhou \\
       &  Hefei \\
       & Hohhot \\
       &  Xinxiang \\
       & Huludao \\
       & Jiaozuo \\
       & Qingdao \\
       &  Jinchang \\
       & Jinzhou \\
       & Kunming \\
       &  Laiwu \\
       & Lanzhou \\
       & Leshan \\
       & Lianyungang \\
       & Liaocheng \\
       &  Linyi \\
       & Luoyang \\
       & Ma'anshan \\
       & $\ddot{U}$r$\ddot{u}$mqi \\
       &  Nanchang \\
       & Nanjing \\
       & Nanning \\
       & Tianjin \\
       & Panzhihua \\
       & Huai'an \\
       & Qinhuangdao \\
       & Qujing \\
       &  Shanghai \\
       &  Shenyang \\
       & Shijiazhuang \\
       & Gongzhuling \\
       & Taiyuan \\
       &  Tangshan \\
       & Dashiqiao \\
       &  Wuhan \\
       &  Xi'an \\
       &  Yinchuan \\
       & Zhengzhou \\
       & Qiqihar \\
       &  Zunyi \\
       &  Shizuishan \\
       & Sanmenxia \\
       & Nanyang \\
       & Yantai \\
       &   Zibo \\
       &   Yuxi \\
       & Xiamen \\
       &  Xiangtan \\
       & Zigong \\
       &  Tongling \\
\cline{1-1}
    IN & Jamshedpur \\
\cline{1-1}
    JP & Kamisu \\
\cline{1-1}
    KR & Pohang-si \\
\hline
 \: & \textbf{\textit{China low Fitness}}\\
\hline
    CN & Aksu \\
       & Altay \\
       &    Ankang \\
       & Baise \\
       &   Baishan \\
       &   Baoshan \\
       &  Bayannur \\
       &    Beihai \\
       & Benxi \\
       &    Bozhou \\
       & Changji/Sanji \\
       &  Chaoyang \\
       &   Chengde \\
       &   Chifeng \\
       &   Lucheng \\
       &   Chuzhou \\
       &  Dali \\
       &    Dazhou \\
       &  Huangshi \\
       & Duyun \\
       & Enshi \\
       &   Ganzhou \\
       &  Gongnong \\
       &   Guigang \\
       &    Guyuan \\
       &  Hanzhong \\
       &  Hebi \\
       & J$\bar{i}$nch$\acute{e}$ngji$\bar{a}$ng \\
       &    Hegang \\
       &  Hengshui \\
       &  Hengyang \\
       &  Heze \\
       &    Hezhou \\
       &   Huaihua \\
       & Hulunbuir \\
       &   Jiamusi \\
       & Jiayuguan \\
       & Jilin \\
       &  Jingdezhen \\
       &  Jinghong \\
       &  Jinzhong \\
       &    Jishou \\
       &   Jiuquan \\
       &   Kaifeng \\
       & Kaili \\
       &   Kashgar \\
       & Hotan \\
       &  Tianshan \\
       & Lhasa \\
       &  Liaoyang \\
       &    Linxia \\
       &   Liuzhou \\
       &   Longnan \\
       &   Louxing \\
       &   Meizhou \\
       &    Mengzi \\
       &  Nanchong \\
       & Ordos \\
       &    Panjin \\
       &  Songyuan \\
       &  Qingyang \\
       &   Qinzhou \\
       &  Shangluo \\
       &  Shangqiu \\
       &  Shaoyang \\
       &    Suihua \\
       &   Suining \\
       &    Turpan \\
       &    Weinan \\
       &   Wenshan \\
       & Wuwei \\
       &   Wuzhong \\
       &  Xianning \\
       &   Xichang \\
       &    Xingyi \\
       &    Xining \\
       & Ya'an \\
       & Yanji \\
       & Yibin \\
       &   Yingtan \\
       &  Yongzhou \\
       &  Yuncheng \\
       & Zhangjiajie \\
       &   Zhangye \\
       &  Xilinhot \\
       &   Sanming \\
       &   Meishan \\
       & Yining/Qulja \\
       &   Ulanqab \\
       &   Xinzhou \\
       &  Zhongwei \\
       & Mudanjiang \\
       & Zhanjiang \\
       & Zhangzhou \\
       &  Neijiang \\
       &  Zhaotong \\
       & Blagoveshchensk \\
       &    Shiyan \\
       &  Shaoguan \\
       &  Shangrao \\
       &  Shuozhou \\
       & Xinyu \\
       &    Panshi \\
       & Yulin \\
       &    Ziyang \\
       & Pingdingshan \\
       &    Rizhao \\
       & Zhumadian \\
       &   Zhoukou \\
       &   Xuchang \\
\hline
     \: &   \textbf{\textit{Emerging economies}} \\
\hline 
    AU & Hobart \\
\cline{1-1}
BR & Recife \\
  &              Campinas \\
  &             Fortaleza \\
  &         Florian$\acute{o}$polis \\
  &        S$\tilde{a}$oCarlos\\
  &        Ribeirao Preto \\
  &             Ceil$\hat{a}$ndia \\
  &           Jo$\tilde{a}$o Pessoa \\
  &              Londrina \\
  &                Goiana \\
  &               Maring$\acute{a}$ \\
  &         Montes Claros \\
  &               Pelotas \\
  &                Vi\c cosa \\
  &            Uberl$\hat{a}$ndia \\
  &  Vit$\acute{o}$ria da Conquista \\
\cline{1-1}
CA &                Sarnia \\
\cline{1-1}
CH &                 Basel \\
\cline{1-1}
   CL &            Concepci$\acute{o}$n\ \\
\cline{1-1}
 CN &                 Sanya \\
   &             Pingliang \\
\cline{1-1}
   CZ &        Hradec Kr$\acute{a}$lov$\acute{e}$ \\
\cline{1-1}
   DE &             Darmstadt \\
 &     Frankfurt am Main \\
    &              Mannheim \\
\cline{1-1}
    ES &                Alacant \\
     &                 C$\acute{a}$diz \\
   &               C$\acute{o}$rdoba \\
     &       Vitoria-Gasteiz \\
\cline{1-1}
  FR &               Dunkirk \\
\cline{1-1}
    GR &             Heraklion \\
    &                Patras \\
\cline{1-1}
HU &              Debrecen \\
     &                  P$\acute{e}$cs \\
   &                Szeged \\
\cline{1-1}
IN &               Lucknow \\
  &                 Surat \\
   &               Madurai \\
  &    Thiruvananthapuram \\
   &                Mysuru \\
  &              Vadodara \\
\cline{1-1}
   IS &             Reykjavik \\
\cline{1-1}
  IT &               Palermo \\
 &              Brindisi \\
  &               Sassari \\
\cline{1-1}
JP &                Shunan \\
   &                   Ube \\
\cline{1-1}
MX &     Cadereyta Jim$\acute{e}$nez \\
 &            Cuernavaca \\
  &            Guanajuato \\
 &            Hermosillo \\
  &              Saltillo \\
\cline{1-1}
NL &                Arnhem \\
  &            Maastricht \\
 &               Heerlen \\
\cline{1-1}
NZ &               Dunedin \\
\cline{1-1}
 PL &                Lublin \\
\cline{1-1}
PT &                 Braga \\
   &               Coimbra \\
\cline{1-1}
RU &               Saratov \\
\cline{1-1}
US &               Saginaw \\
 &             Davenport \\
\cline{1-1}
  ZA &          Bloemfontein \\
\hline
         \: & \textbf{\textit{High Tech}}\\
\hline 
     AU &             Melbourne \\
   &              Brisbane \\
\cline{1-1}
     CA &               Toronto \\
        &               Detroit \\
        &                Ottawa \\
        &                London \\
\cline{1-1}
     ES &                Girona \\
        &               Granada \\
\cline{1-1}
     FR &             Marseille \\
        &                  Nice \\
        &           Montpellier \\
\cline{1-1}
     GB &               Bristol \\
        &                London \\
        &             Cambridge \\
        &         Middlesbrough \\
        &               Ipswich \\
        &              Plymouth \\
\cline{1-1}
     IN &             Bengaluru \\
\cline{1-1}
     MX &           Mexico City \\
\cline{1-1}
     NL &             Eindhoven \\
\cline{1-1}
     US &              San Jose \\
        &       Washington D.C. \\
        &             Baltimore \\
        &           San Antonio \\
        &               Atlanta \\
        &                Austin \\
        &              New York \\
        &           Albuquerque \\
        &                 Miami \\
        &            Louisville \\
        &          Philadelphia \\
        &         Oklahoma City \\
        &                 Nampa \\
        &            Charleston \\
        &        Virginia Beach \\
        &                Dallas \\
        &                 Ogden \\
        &                 Omaha \\
        &             Lexington \\
        &           Tallahassee \\
        &                 Tampa \\
        &              Honolulu \\
        &           Los Angeles \\
        &               Phoenix \\
        &              Monterey \\
        &                Tucson \\
        &           Kansas City \\
        &               Tijuana \\
        &            Sacramento \\
        &              Palm Bay \\
        &               Seattle \\
        &               Orlando \\
        &             Bradenton \\
        &               Chicago \\
\cline{1-1}
     ZA &             Cape Town \\
\hline
     \: & \textbf{\textit{Korea \& Japan}}\\
\hline
CA &                Guelph \\
\cline{1-1}
     CN &                Anyang \\
        &                 Jinan \\
\cline{1-1}
     EE &               Tallinn \\
\cline{1-1}
     ES &               Seville \\
        &                Murcia \\
        &     Palma de Mallorca \\
        &               Almeria \\
\cline{1-1}
     FR &           La Rochelle \\
        &       Aix-en-Provence \\
        &               Avignon \\
        &      Clermont-Ferrand \\
\cline{1-1}
     JP &             Tomakomai \\
        &             Kagoshima \\
        &                Aomori \\
        &             Asahikawa \\
        &                  Naha \\
        &               Fukuoka \\
        &               Sapporo \\
        &              Shizuoka \\
        &             Hachinohe \\
        &              Hakodate \\
        &                 K$\bar{o}$chi \\
        &                Yonago \\
        &                 Izumo \\
        &                Kitami \\
        &               Kushiro \\
        &               Obihiro \\
        &              Kan'onji \\
        &               Morioka \\
        &              Hirosaki \\
        &            Miyakonojo \\
        &               Niigata \\
        &                 $\bar{O}$mura \\
        &                 Izumimachikurosuno \\
        &                  $\bar{O}$ita \\
\cline{1-1}
     KR &                 Seoul \\
        &            Cheonan-si \\
        &                 Busan \\
        &           Cheongju-si \\
        &          Chuncheon-si \\
        &            Chungju-si \\
        &                 Daegu \\
        &               Gwangju \\
        &              Iksan-si \\
        &                Jeonju \\
        &              Mokpo-si \\
        &              Wonju-si \\
        &             Gunsan-si \\
        &            Sora-myeon \\
\cline{1-1}
     NL &                 Breda \\
\cline{1-1}
     NZ &              Hamilton \\
\hline
          \: & \textbf{\textit{Western}}\\
\hline 
AU &            Gold Coast \\
\cline{1-1}
     CA &            Abbotsford \\
        &                Barrie \\
        &              Red Deer \\
        &              Winnipeg \\
        &              Kingston \\
\cline{1-1}
     ES &              A Coru$\tilde{n}$a \\
        &               Logro$\tilde{n}$o \\
        &              Zaragoza \\
        &                 Gij$\acute{o}$n \\
        &                Lleida \\
\cline{1-1}
     FI &             Jyv$\ddot{a}$skyl$\ddot{a}$ \\
        &                Kuopio \\
\cline{1-1}
     FR &               Ajaccio \\
        &             Perpignan \\
        &                  Metz \\
        &                Angers \\
        &                Troyes \\
        &               Le Mans \\
        &               Lorient \\
        &                Toulon \\
        &               B$\acute{e}$ziers \\
        &             Cherbourg \\
        &          Saint-Brieuc \\
\cline{1-1}
     GB &              Brighton \\
        &               Ashford \\
        &       Southend-on-Sea \\
        &             Blackburn \\
        &             Blackpool \\
        &           Bournemouth \\
        &              Bridgend \\
        &               Norwich \\
        &               Cardiff \\
        &              Carlisle \\
        &               Swansea \\
        &                 Crewe \\
        &               Preston \\
        &            Colchester \\
        &                 Corby \\
        &                Exeter \\
        &            Portsmouth \\
        &        Stoke-on-Trent \\
        &            Eastbourne \\
        &   Newcastle upon Tyne \\
        &        Great Yarmouth \\
        &           St Leonards \\
        &          Peterborough \\
        &             Lancaster \\
        &               Lincoln \\
        &                 Luton \\
        &               Grimsby \\
        &            Scunthorpe \\
        &               Fazeley \\
        &              Redditch \\
        &               Taunton \\
        &               Torquay \\
        & Royal Tunbridge Wells \\
        &            Chelmsford \\
        &           Dunfermline \\
        &               Chester \\
        &               Glasgow \\
        &             Liverpool \\
\cline{1-1}
     IT &       Bolzano - Bozen \\
\cline{1-1}
     NL &                 Hoorn \\
\cline{1-1}
     NZ &              Auckland \\
\cline{1-1}
     TR &               Kapakl{\i} \\
\cline{1-1}
     US &             Anchorage \\
        &               Killeen \\
        &              Waterloo \\
        &              Bismarck \\
        &                 Chico \\
        &            Shreveport \\
        &          Lake Charles \\
        &            Cape Coral \\
        &                Naples \\
        &            Fort Smith \\
        &         Winston-Salem \\
        &               Dubuque \\
        &            Palm Coast \\
        &                 Flint \\
        &           Port Arthur \\
        &           Spring Hill \\
        &              Kankakee \\
        &             La Crosse \\
        &               Lubbock \\
        &          Wilkes-Barre \\
        &                Madera \\
        &        Port St. Lucie \\
        &                Boston \\
        &      Lake Havasu City \\
        &        Corpus Christi \\
        &               Navarre \\
        &              Brockton \\
        &              Amarillo \\
        &                 Tracy \\
        &                Topeka \\
        &               Modesto \\
        &               Abilene \\
        &               Visalia \\
        &                Yakima \\
        &                  Yuma \\
        &               Deltona \\
        &          Murfreesboro \\
        &          Jacksonville \\
        &          Nuevo Laredo \\
        &           Saint Cloud \\
        &         Wichita Falls \\
        &              Savannah \\
        &        Salt Lake City \\
\cline{1-1}
     ZA &                Durban \\
\hline
      \: & \textbf{\textit{Western Cars}}\\
\hline 
 DE &             Stuttgart \\
        &            Ingolstadt \\
\cline{1-1}
     FR &           Montb$\acute{e}$liard \\
\cline{1-1}
     GB &            Birmingham \\
        &              Coventry \\
        &               Warwick \\
\cline{1-1}
     IT &                 Turin \\
        &    Reggio nell'Emilia \\
\cline{1-1}
     JP &                Nagoya \\
        &             Hamamatsu \\
\cline{1-1}
     US &               Detroit \\
        &                Peoria \\
        &            Janesville \\
        &             Ann Arbor \\
\hline
\end{supertabular}
\end{multicols*}

\onecolumn
\fontsize{12}{10}\selectfont

\section*{References}
[1] Kummu, M., Taka, M. \& Guillaume, J. H. Gridded global datasets for gross domestic product and human development index over 1990-2015. Sci. data 5, 1-15 (2018)